\documentclass[nofootinbib,pra,twocolumn,showpacs,superscriptaddress,notitlepage,groupedaddress,amsmath]{revtex4-2} 
\usepackage[colorlinks=true,citecolor=blue,linkcolor=blue,urlcolor=blue]{hyperref}

\usepackage{amsmath,amssymb,bm,color}

\usepackage{framed}
\definecolor{shadecolor}{gray}{0.95}
\usepackage{amsthm}

\usepackage{tikz}
\usetikzlibrary{external}
\usetikzlibrary{decorations.pathreplacing}
\usetikzlibrary{decorations.text}

\definecolor{newblue}{RGB}{112,178,255}
\definecolor{neworange}{RGB}{255,204,112}
\definecolor{blue2}{RGB}{120,0,255}
\definecolor{red2}{RGB}{255,0,120}
\definecolor{green2}{RGB}{0,130,130}

\definecolor{darkred}{RGB}{245,186,183}
\definecolor{lightred}{RGB}{249,217,215}

\def\ket#1{\mathinner{|{#1}\rangle}}
\def\Ket#1{\mathinner{\left|{#1}\right\rangle}}
\def\Bra#1{\mathinner{\left\langle{#1}\right|}}
\def\bs#1{\boldsymbol{#1}}
\def\inner#1{\mathinner{\langle{#1}\rangle}}

\def\ZZ{\mathbb Z}
\def\cX{\mathcal X}
\def\cZ{\mathcal Z}
\def\cC{\mathcal C}
\usetikzlibrary{arrows,calc,external,shapes.geometric}

\tikzset{
	>=stealth',
	help lines/.style={dashed, thick},
	important line/.style={thick},
	connection/.style={thick, dotted},
}

\tikzstyle{A}=[circle,draw=red!50,fill=red!20,thick]
\tikzstyle{R}=[circle,draw=blue!50,fill=blue!20,thick]
\tikzstyle{U}=[circle,draw=green!50,fill=green!20,thick]
\tikzstyle{V}=[circle,draw=orange!50,fill=orange!20,thick]

\begin{document}


\title{Efficiently preparing Schr\"odinger's cat, fractons and non-Abelian topological order\\in quantum devices}

\author{Ruben Verresen}

\author{Nathanan Tantivasadakarn}

\author{Ashvin Vishwanath}

\address{Department of Physics, Harvard University, Cambridge, MA 02138, USA}

\date{\today}

\maketitle

\textbf{Long-range entangled quantum states---like cat states and topological order \cite{Wenbook}---are key for quantum metrology and information purposes \cite{Dennis02,Fowler_2012}, but they cannot be prepared by any scalable unitary process \cite{BravyiHastingsVerstraete06,Hastings2010}. Intriguingly, using measurements as an additional ingredient could circumvent such no-go theorems \cite{Gottesman97,Briegel01,Raussendorf05,Bolt16,Piroli21}.
However, efficient schemes are known for only a limited class of long-range entangled states, and their implementation on existing quantum devices via a sequence of gates and measurements is hampered by high overheads.
Here we resolve these problems, proposing how to scalably prepare a broad range of long-range entangled states with the use of existing experimental platforms.
Our two-step process finds an ideal implementation in Rydberg atom arrays \cite{Jaksch00,Lukin01,Urban09,Gaetan09,Isenhower10,Wilk10,Comparat10,Weimer10,Bernien17,deLeseleuc19,Omran19,Browaeys20,Semeghini21,Ebadi21,Scholl21,SCholl21a}, only requiring time-evolution under the intrinsic atomic interactions, followed by measuring a single sublattice (by using, e.g., two atom species \cite{Liu18,Brooks21,Singh21,Zhang21}). Remarkably, this protocol can prepare the 1D Greenberger-Horne-Zeilinger (GHZ) `cat' state and 2D toric code \cite{Kitaev_2003} with fidelity per site exceeding $0.9999$, and a 3D fracton state \cite{Haah2011,Yoshida13,Vijay16}
with fidelity $\gtrapprox 0.998$. In light of recent experiments showcasing 3D Rydberg atom arrays \cite{Barredo18}, this paves the way to the first experimental realization of fracton order. While the above examples are based on efficiently preparing and measuring cluster states \cite{Briegel01}, we also propose a multi-step procedure to create $S_3$ and $D_4$ non-Abelian topological order in Rydberg atom arrays and other quantum devices---offering a route towards universal topological quantum computation \cite{Mochon04}.
}

The defining property of long-range entangled states is that they cannot be prepared efficiently by unitary processes, that is, prepared within a fixed  time window independent of system size. Remarkably, measurements can 
circumvent this issue in quantum devices. In particular, certain long-range entangled states can in principle be prepared in a fixed amount of time by first acting with simple quantum gates (typically controlled-$Z$ or Ising)---preparing the so-called cluster state \cite{Briegel01}. Subsequently measuring a subset of quantum bits leaves the remainder in a long-range entangled state \cite{Briegel01,Raussendorf05,Bolt16,Piroli21}.

\begin{figure*}
    \centering
\begin{tikzpicture}
\node at (-0.1,0.1) {\includegraphics[scale=1.1]{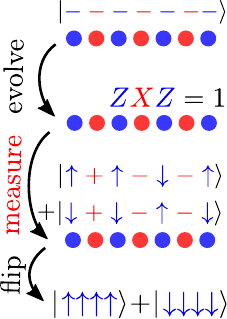}};
\node at (4.1,0) {\includegraphics[scale=0.33]{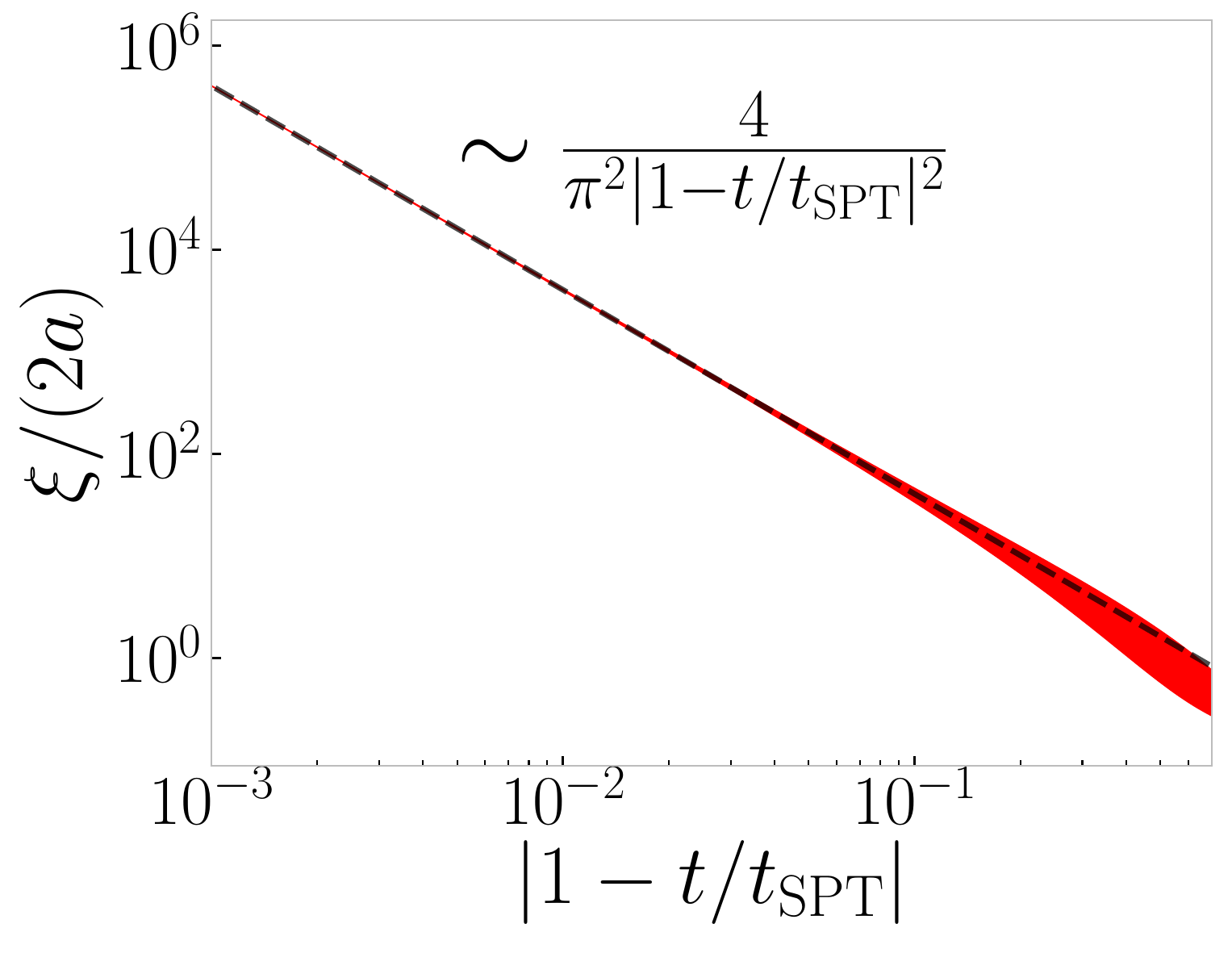}};
\node at (10,0) {\includegraphics[scale=0.33]{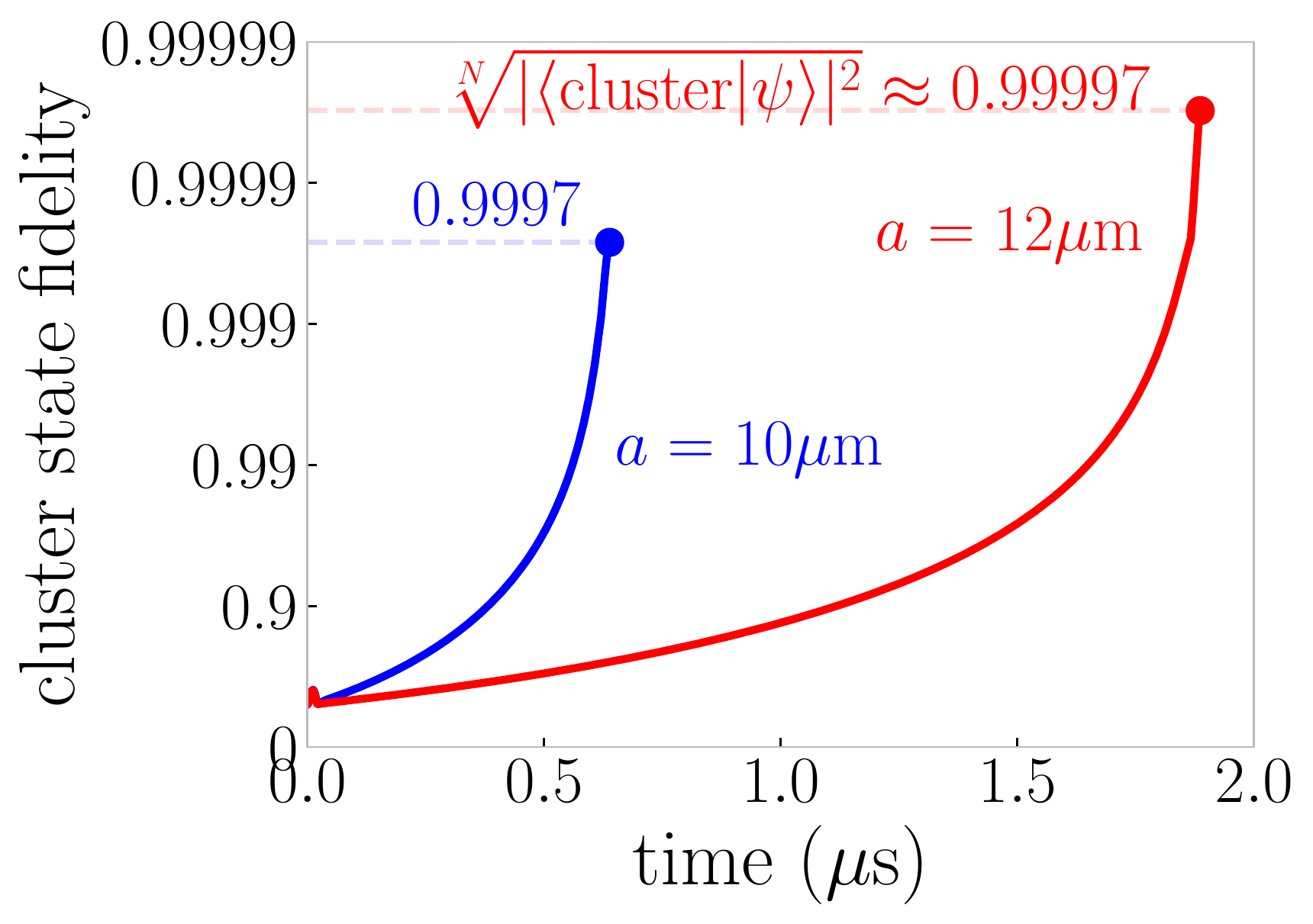}};
\node at (14.5,-0.1)
{\includegraphics[scale=1.1]{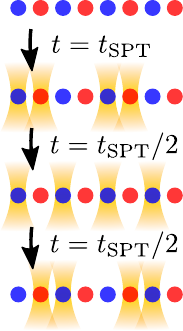}};
\node at (-1.4,1.7) {(a)};
\node at (1.7,1.7) {(b)};
\node at (7,1.7) {(c)};
\node at (13.2,1.7) {(d)};
\end{tikzpicture}
    \caption{\textbf{Fast preparation of GHZ state by measuring 1D cluster state.} (a) The protocol starts with a product state in the $X$-basis which we let time-evolve under an Ising interaction. After a time $t=t_\textrm{SPT}$, this produces the cluster state. Measuring the red sites in the $X$-basis leads to a cat state. Using the information of the measurement outcomes, one can flip the appropriate spins to obtain the GHZ state on the blue sites \cite{Briegel01}. (b) If we do not time-evolve with exactly $t = t_\textrm{SPT}$, the post-measurement state is only an approximate cat state. Nevertheless, the resulting correlation length (within which there is long-range order) is very large even for moderate time deviations. (c) A tensor-network simulation of the protocol for Rydberg atoms interacting with a $1/r^6$ van der Waals interaction. We simulate this for Rb $70S_{1/2}$ with two different lattice spacings, which give different results due to incorporating the fact that the initial product state preparation pulses are not instantaneous (see \hyperref[methods]{Methods}). (d) Contributions from longer-range van der Waals interactions can be systematically suppressed by interspersing time-evolution with $X$-pulses (orange) on particular sublattices (see \hyperref[methods]{Methods}). This particular example cancels out couplings at distance $r=2a$, although this was not used to achieve the results in (c). \label{fig:1D}}
\end{figure*}
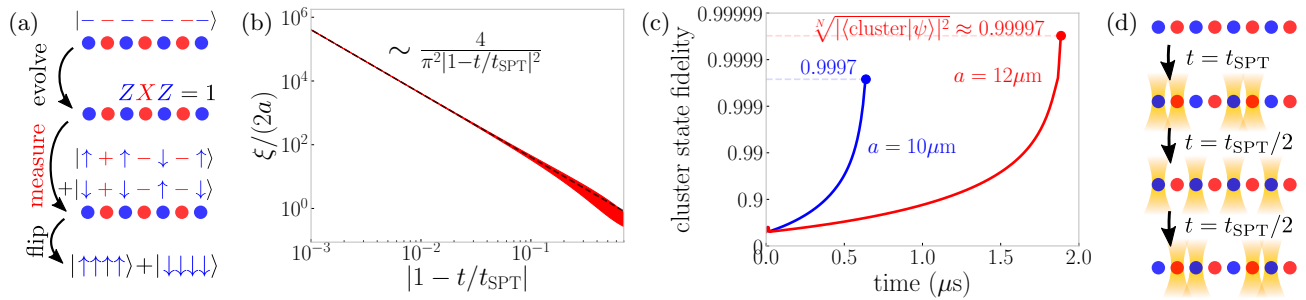

Some well-known examples are the cluster state on a 1D chain or the 2D Lieb lattice, where single-site measurements transform these into the GHZ `cat' state \cite{Briegel01} or the toric code \cite{Raussendorf05}, respectively. However, even these known cases have resisted experimental implementation. Indeed, the infrastructure necessary to apply such quantum gates can require a high overhead, and setting aside a fraction of the qubits to be measured is costly in emerging quantum computers---with recent realizations of the toric code phase \cite{Kitaev_2003} on quantum devices opting for a unitary protocol \cite{Satzinger21,Semeghini21}. Moreover, going beyond such known examples, an important open question is what other exotic state with potentially valuable properties could be prepared within a measurement-based protocol.

Here, we address both these  issues. In terms of novel examples, we present 3D cluster states where single-site measurements produce the elusive fracton order---an intriguing family of 3D states which have resisted experimental detection and hold promise as optimal quantum hard drives \cite{Haah2011,Shirley18}. Moreover, going beyond the cluster state framework \cite{Brennen09,Brell15,Bolt16}, we reveal how non-Abelian topological order can be obtained via \emph{multiple} alternations between gates and measurements, establishing the first proposal for efficiently obtaining such an exotic state in any platform. The second issue we successfully address is that of implementability: all of the aforementioned examples---old and new---can be realized using \emph{existing} Rydberg atom technology. In particular, we let the entangling operations naturally emerge from the time-evolution under the intrinsic interactions between Rydberg atoms, rather than constructed piecemeal via the gates of a quantum computer, thereby trading in the overhead of the latter for scalability.

Rydberg atom arrays \cite{Jaksch00,Lukin01,Urban09,Gaetan09,Isenhower10,Wilk10,Comparat10,Browaeys20} offer an ideal platform satisfying the two requirements of Ising interactions and high-resolution measurements. These are tuneable lattices of atoms trapped by optical tweezers, each serving as an effective qubit \cite{Jones07} consisting of its ground state ($n=0$) and a particular Rydberg excited state ($n=1$). When two atoms are simultaneously excited, they experience a van der Waals interaction $V(r) \sim 1/r^6$ \cite{Sibalic18}, giving rise to a Hamiltonian of the form \cite{Robicheaux05}:
\begin{equation}
H = \sum_{\bm i} \frac{\Omega_{\bm i}}{2} X_{\bm i} - \sum_{\bm i} \delta_{\bm i}n_{\bm i} + \frac{1}{2} \sum_{\bm i \neq \bm j} V_{\bm{ij}} n_{\bm i} n_{\bm j}, \label{eq:Hryd}
\end{equation}
where $X,Y,Z$ denote the Pauli matrices. Here $\Omega_{\bm i}$ and $\delta_{\bm i}$ are tuneable, and by shifting the latter, we obtain an explicit Ising interaction $\frac{1}{2} \sum_{\bm i \neq \bm j} J_{\bm{ij}} Z_{\bm i} Z_{\bm j}$ with $J_{\bm{ij}} = \frac{1}{4} V_{\bm{ij}}$, which will play the role of our `entangling gates'. Recent years have seen a rapid growth in available system sizes, exploring the many-body quantum physics of Eq.~\eqref{eq:Hryd} in 1D and 2D arrays of many hundreds of atoms \cite{Bernien17,deLeseleuc19,Omran19,Semeghini21,Ebadi21,Scholl21,SCholl21a,Ebadi21}, and even demonstrating 3D arrays \cite{Barredo18}. Confirming the platform's promise for realizing exotic states, Ref.~\onlinecite{Semeghini21} recently established the onset of toric code topological order based on a proposal in Ref.~\onlinecite{VerresenLukinVishwanath21}.

In the simplest instances, our state preparation procedure is a two-step protocol of time-evolution and measurement. At the end of the first step, we obtain high-quality cluster states despite the longer-range van der Waals corrections. This is due to a judicious choice of lattices, as well as reinterpreting the power of cluster states as being a property of their symmetry-protected topological (SPT) \cite{Gu09,Son11} phase. While cluster states are already of interest (serving as resources for measurement-based quantum computation \cite{RaussendorfBriegel2001}), to produce long-range entanglement we need to be able to measure only a subsystem of its atoms. Recent breakthrough experiments have established that this is possible in dual-species arrays \cite{Beterov15,Samboy17,Otto20,Liu18,Brooks21,Singh21,Zhang21}, where one species is transparent to measurements of the other. This way, we produce the 1D GHZ, the 2D toric code and 3D fracton states.

Non-Abelian $D_4$ or $S_3$ topological order can be obtained by using a four-step protocol where one time-evolves and measures twice, creating the color code \cite{Bombin06} or $\mathbb Z_3$ toric code \cite{Kitaev_2003} as an intermediate state. This can be conceptually interpreted as sequentially gauging certain symmetries, thereby expanding the possible states of matter which can be efficiently prepared by finite-time evolution and measurement, the theoretical ramifications of which we explore further in a companion work \cite{framework}.

\section{The 1D cluster and GHZ states \label{sec:1D}}

To describe how to efficiently prepare a Schr\"odinger cat state, let us first consider the idealized setting of nearest-neighbor Ising interactions, after which we discuss the effect of imperfections. Starting with a product state in the Pauli-$X$ basis, we time-evolve a chain of $N$ qubits:
\begin{equation}
|\psi(t) \rangle = e^{-i t J \sum_n Z_n Z_{n+1}} |-\rangle^{\otimes N}. \label{eq:ideal}
\end{equation}
After a time $t_\textrm{SPT} := \frac{\pi}{4|J|}$, the resulting wave function forms the so-called 1D cluster state \cite{Briegel01}. Its defining characteristic is:
\begin{equation}
Z_{n-1} X_n Z_{n+1} |\psi(t_\textrm{SPT})\rangle = |\psi(t_\textrm{SPT})\rangle. \label{eq:cluster_stabilizer}
\end{equation}
This implies that the entangled state has the remarkable property that measuring, say, every even site in the $X$-basis leaves the remaining qubits in a Schr\"odinger cat state \cite{Briegel01}. Indeed, replacing $X_{2n} \to \pm 1$ by the measurement outcome, the post-measurement state satisfies $Z_{2n-1} Z_{2n+1}|\psi_\textrm{out}\rangle = \pm |\psi_\textrm{out}\rangle$. In addition, one can show that the state is symmetric, i.e., $\langle Z_{2n-1} \rangle = 0$ (see \hyperref[methods]{Methods}). The only state satisfying these properties, is the (disordered) GHZ state, a Schr\"odinger cat state where neighboring spins on the odd sublattice are perfectly aligned or anti-aligned depending on the measurement outcome of the intervening site, as illustrated in Fig.~\ref{fig:1D}(a). Furthermore, since this information is available, one can always apply single-site spin flips to bring it to the ideal GHZ state:
\begin{equation}
|\textrm{GHZ}\rangle = \frac{|\uparrow\uparrow \cdots \uparrow \rangle + |\downarrow \downarrow \cdots \downarrow \rangle}{\sqrt{2}},
\end{equation}
although it is a Schr\"odinger cat state independent of this procedure.

To make this procedure practical, we need to understand its sensitivity to deviations from the above idealization, such as evolving with additional couplings or for the incorrect time. We argue that the ability to create a cat state via measurement is a stable property of the phase of matter that the cluster state belongs to. More precisely, it is known that in addition to the global Ising symmetry $\prod_n X_n$, the cluster state is symmetric under spin-flips on even and odd sites, separately; deformations of the cluster state that preserve this symmetry define its so-called SPT phase \cite{Gu09,Son11}.

As a first indication of the importance of this symmetry, we explore what happens when we slightly break it by time-evolving for the wrong duration. If $t$ is not a multiple of $t_\textrm{SPT}$, the state does not enjoy the aforementioned sublattice symmetry. Indeed, a straightforward computation shows that (ignoring boundary effects and taking $|1-t/t_\textrm{SPT}|<1/2$):
\begin{equation}
\langle \psi(t)| \prod_n X_{2n} |\psi(t) \rangle = \sin(2Jt)^{N} \sim e^{-N/\xi_s(t)}. \label{eq:symbreak}
\end{equation}
This defines a length scale $\xi_s(t)$ within which the state looks symmetric, with $\xi_s(t_\textrm{SPT})=\infty$. In line with the idea that the symmetry is the determining factor, we surmise that after measuring every other site, we obtain an imperfect cat state with correlation length $\xi_s(t)$. Indeed, we have confirmed by a full solution of the model that the resulting state has long-range order for $\langle Z_{2m-1} Z_{2n-1} \rangle$ within a correlation length $\xi$ closely matching $\xi_s$, beyond which it decays to zero \cite{SM}. This is plotted in Fig.~\ref{fig:1D}(b), where we see that $\xi$ grows rapidly as we approach $t \to t_\textrm{SPT}$. E.g., a $0.1\%$ deviation results in long-range order over nearly half a million atoms; a $1\%$ error still gives $\xi/(2a) \approx 4000$ (the original lattice having spacing $a$). Hence, even in the presence of timing imperfections, this procedure can produce large
cat states in a finite amount of time.

A chain of Rydberg atoms gives a good approximation to Eq.~\eqref{eq:ideal} with $J = \frac{V(a)}{4}$, with additional longer-range corrections. In case we use a single species of Rydberg atoms, the first correction occurs at distance $r=2a$ with a van der Waals coupling $V(2a) = V(a)/2^6$, giving an additional evolution by $e^{-\frac{1}{2^6} itJ \sum_n Z_n Z_{n+2}}$. However, one can show that since this Ising coupling preserves the spin-flip symmetry on even and odd sublattices, measuring the even sites still leads to a cat state with an infinite correlation length (see \hyperref[methods]{Methods}). It is thus only the third-nearest-neighbor correction which will lead to an imperfect cat state. Similar to Eq.~\eqref{eq:symbreak}, we find that this leads to a correlation length
\begin{equation}
\frac{\xi}{2a} = \frac{1}{2 \left|\ln \cos\left(\pi/(2\times 3^6)\right)\right|} \approx 2 \times 10^5. 
\label{eq:1D_xi}
\end{equation}
We thus obtain a near-perfect cat state.

In addition to knowing the size of the cat state that we prepare, we can ask how close we are to the fixed-point GHZ limit. A useful fact is that, to a very high degree, we can equate the GHZ-fidelity of the final state with the cluster-state-fidelity of the pre-measurement state (see \hyperref[methods]{Methods}). To characterize the latter, we can use the fact that the expectation value of the stabilizer $Z_{n-1} X_n Z_{n+1}$ gives a lower bound on the fidelity (see \hyperref[methods]{Methods}). At first sight, this expectation value is affected by the second-nearest neighbor coupling at $r=2a$. However, in this 1D case, we can show that its effect is to simply tilt the Ising order parameter away from $Z_n$, which can be easily corrected by a single-site rotation at the end of the time-evolution. In conclusion, even for a single species the dominant correction is at $r=3a$, leading to a cluster stabilizer
\begin{equation}
\langle Z_{n-1} X_n Z_{n+1} \rangle \approx \cos^2 \left( \frac{\pi}{2 \times 3^6} \right) \approx 0.999995.
\end{equation}
This suggests a fidelity per site of
\begin{align}
\sqrt[N]{|\langle \textrm{cluster}|\psi(t_\textrm{SPT})\rangle|^2} &\gtrapprox \frac{1+\langle Z_{n-1} X_n Z_{n+1}\rangle}{2} \\
 & \approx 0.9999975. \label{eq:ideal}
\end{align}
This is such a staggeringly high fidelity, that one might worry that at this point other imperfections---which we might usually reasonably ignore---become dominant. One such imperfection is the fact that the initial product state is prepared by a pulse, which has a finite time on the order of tens of nanoseconds. We have performed a full tensor network simulation of our state preparation for the case of Rb $70S_{1/2}$, incorporating this imperfection. As we see in Fig.~\ref{fig:1D}(c), we can reach very high fidelities. The fact that the resulting fidelity is one order of magnitude worse than Eq.~\eqref{eq:ideal} is due to the finite pulse time. Indeed, we have confirmed that artificially reducing the pulse time recovers Eq.~\eqref{eq:ideal} (see \hyperref[methods]{Methods}).

\section{Toric and color code \label{sec:toric}}

\begin{figure}
    \centering
    \begin{tikzpicture}
    \node at (0,0) {\includegraphics[scale=0.55]{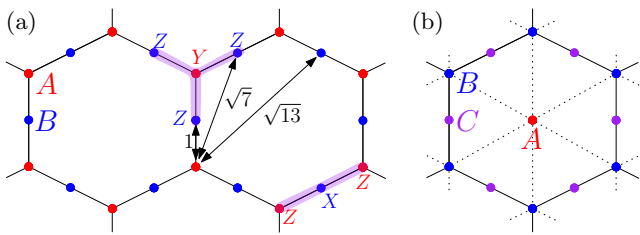}};
    \node at (-4,1.3) {(a)};
    \node at (1.4,1.3) {(b)};
    \end{tikzpicture}
    \caption{\textbf{From toric and color code to non-Abelian topological order.} (a) We place Rydberg atoms on the vertices (red) and bonds (blue) of the honeycomb lattice. Initializing into the product state $|-\rangle^{\otimes N}$ and time-evolving for $t=\pi/V(a)$, we obtain the cluster state whose two types of stabilizers are depicted. Measuring the Rydberg atoms on the red sites in the $Y$-basis produces the toric code state on the blue sites up to known single-site spin flips.
    Corrections to topological order are given by longer-range van der Waals interactions connecting the red and blue sublattices; the leading correction has a pre-factor $1/\sqrt{7}^6 \approx 0.003$.
    (b) If one instead puts atoms on the centers (red) and vertices (blue) of the hexagons, then time-evolution and measurement of the red sites produces the color code. Subsequently loading atoms on the links (purple) can produce $D_4$ non-Abelian topological order after another round of time-evolution and measuring the blue sites (see Section~\ref{sec:nonabelian}).}
    \label{fig:TC}
\end{figure}

Having established cat states in 1D, we now turn to 2D topological order. We consider qubits on the vertices ($A$ sublattice) and bonds ($B$ sublattice) of the honeycomb lattice as shown in Fig.~\ref{fig:TC}(a). Like in the 1D case, we start with a product state $|-\rangle^{\otimes N}$ and time-evolve evolve with a nearest-neighbhor Ising interaction for a time $t_\textrm{SPT}$ to obtain a 2D cluster state \cite{Raussendorf05} which is invariant under the two types of stabilizers shown in Fig.~\ref{fig:TC}(a). In particular, for a given vertex $v$, we have $Y_v \mathcal A_v|\psi(t_\textrm{SPT})\rangle = |\psi(t_\textrm{SPT})\rangle$ where $Y_v$ is the Pauli-$Y$ on that vertex, and $\mathcal A_v = \prod_{b \in v} Z_b$ is a product over the three neighboring bond qubits. Hence, upon measuring the $A$ sublattice in the $Y$-basis, we obtain a state where $\mathcal A_v = \pm 1$ depending on the measurement outcome. Moreover, one can also show that for every hexagonal plaquette we have $\mathcal B_p = \prod_{b \in p} X_b = 1$ (see \hyperref[methods]{Methods}). These are exactly the two stabilizers of the toric code \cite{Kitaev_2003}! The state has $\mathbb Z_2$ topological order for either $\mathcal A_v = \pm 1$. If one wishes to use this state for quantum computation purposes, one can simply store these measurement outcomes and use them in post-processing. Alternatively, one can always bring the state to $\mathcal A_v=+1$ by applying string operators formed by single-site $X$-rotations.

If we implement this using Rydberg atoms, then the time-evolution gets contributions from the longer-range van der Waals interactions. However, like in 1D, we argue that the ability to produce topological order via measurement is a robust property of the entire SPT phase of this 2D state. The practical consequence is that the topological phase is stable to all longer-range Ising couplings which do not couple the $A$ and $B$ sublattice (see \hyperref[methods]{Methods}). The dominant correction is thus at a distance $r=\sqrt{7}a$, shown in Fig.~\ref{fig:TC}. This means that the post-measurement state exhibits topological order only over a finite number of atoms on the $B$ sublattice, which we compute (similar to Eqs.~\eqref{eq:symbreak} and \eqref{eq:1D_xi}) to be
\begin{equation}
\frac{3}{2} \times \frac{1}{6} \times \frac{1}{\big|\ln \cos\big(\pi/\big(2\times \sqrt{7}^6\big)\big) \big|} \approx 2 \times 10^4.
\end{equation}
Here $3/2$ counts the number of $B$ sites per $A$ site, and $6$ represents the number of couplings at distance $r=\sqrt{7}a$. To exemplify the importance of choosing an optimal lattice: for the square and triangular lattices we find $\mathbb Z_2$ topological order for roughly $3000$ and $300$ atoms, respectively.

In addition to achieving the topological phase in a macroscopic system, a more refined question is how close we are to the toric code fixed-point limit. Unlike the above discussion, this depends on the strength of the $AA$ and $BB$ couplings. For concreteness, we consider dual-species arrays, where these intra-species couplings can be made negligible (alternatively, they can be suppressed using local addressing, see \hyperref[methods]{Methods}). The expectation value of the stabilizers is then (see also Fig.~\ref{fig:TC}): 
\begin{align*}
&\left|\langle \mathcal A_v \rangle\right| \approx \cos^6 \left( \frac{\pi}{2\sqrt{7}^6} \right)
\cos^6 \left( \frac{\pi}{2 \sqrt{13}^6} \right)  \cdots \approx 0.99993,\\
&\langle \mathcal B_p \rangle \approx  \cos^{12} \left( \frac{\pi}{2 \sqrt{7}^6} \right) 
 \cos^{12} \left( \frac{\pi}{2  \sqrt{13}^6} \right)  \cdots \approx 0.99987.
\end{align*}
This implies a toric code fidelity $\sqrt[N]{\mathcal F} \gtrapprox 0.9999$, making the state useful for quantum information purposes.
Indeed, it is well-known that terminating the lattice with an alternation of rough and smooth boundaries gives a surface code \cite{Bravyi98}. Applying local $X$- and $Z$-pulses in string patterns creates and moves $e$- and $m$-anyons \cite{Kitaev_2003}, acting as logical operators on the topological qubit. Active error correction \cite{Dennis02,Fowler_2012} requires repeatedly measuring the $\mathcal A_v$ stabilizers, which is achieved by reloading the $A$ sublattice and repeating the preparation procedure. Similarly, the $\mathcal B_p$ stabilizers can be measured by performing a $X \to Z$ rotation and loading atoms at the center of the plaquettes (for which a square lattice is most natural).

This procedure can be generalized to other topological orders. For instance, if the $A$ sublattice lives at the centers of plaquettes of the honeycomb lattice, and $B$ are the vertices (see Fig.~\ref{fig:TC}(b)), then repeating the above procedure produces the color code on the $B$ sublattice, whose stabilizers are $\prod_{v \in p} X_v$ and $\prod_{v\in p} Z_v$ for every plaquette \cite{Bombin06}. For this state---which has significantly more quantum computing power than the toric code---we obtain a fidelity per site $\sqrt[N]{\mathcal F} \gtrapprox 0.997$ \cite{SM}. We can also prepare the 3D toric code---which exhibits more thermal stability \cite{Castelnovo08}---with $\sqrt[N]{\mathcal F} \gtrapprox 0.9998$ on the diamond lattice \cite{SM}.

\begin{figure}
    \centering
    \begin{tikzpicture}
    \node at (0,0) {\includegraphics[scale=0.63]{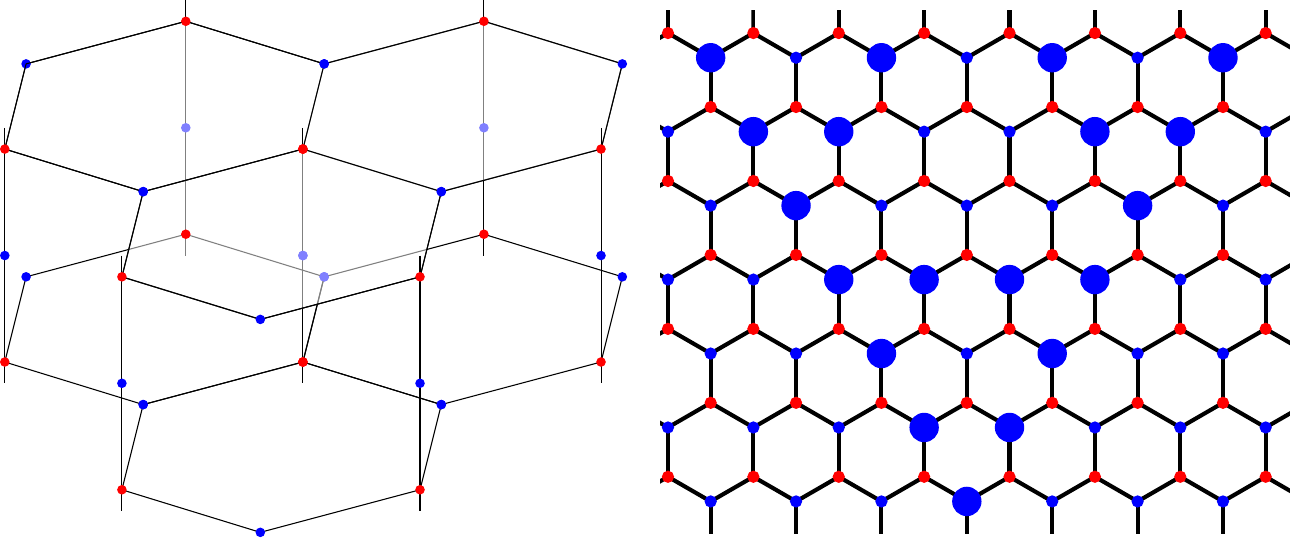}};
    \node at (-4,-1.3) {(a)};
    \node at (-0.3,-1.3) {(b)};
    \end{tikzpicture}
    \caption{\textbf{Realizing a fractal spin liquid with fracton order.} (a) Rydberg atoms are placed at the red and blue sites on the hexagonal prism lattice. After time-evolving under the Rydberg interaction and measuring the red sites, we obtain the ground state of the so-called Sierpinski prism model \cite{Yoshida13} (see \hyperref[methods]{Methods}). (b) Quasiparticles have restricted mobility in the hexagonal layers due to being created at the corners of fractal (Sierpinski) operators.}
    \label{fig:fractaldistance}
\end{figure}

\section{Fracton order \label{sec:fracton}}

To achieve fracton order---a very exotic version of topological order since it cannot be described by a conventional quantum field theory \cite{Shirley18}---we consider qubits on the 3D hexagonal prism lattice shown in Fig.~\ref{fig:fractaldistance}(a), with the red and blue dots defining the $A$ and $B$ sublattices, respectively. As above, time-evolving a product state under a nearest-neighbor Ising interaction prepares the cluster state on this latice. Upon measuring the red sublattice, we obtain a state which can be equated with the fractal spin liquid introduced by Yoshida \cite{Yoshida13} (see \hyperref[methods]{Methods}). This exhibits fracton order. In particular, while particles can move in the $z$-direction by the application of a string operator (as in the toric code), in the hexagonal planes they live at the endpoints of a fractal operator (in particular, the Sierpinski triangle; see Fig.~\ref{fig:fractaldistance}(b)), restricting their mobility.

Unlike the previously discussed cases of the GHZ state or topological order, cluster SPT phases giving rise to fracton order are not protected by a \emph{global} symmetry on the $A$ sublattice, but instead by a \emph{subsystem} or \emph{fractal} one. A practical consequence is that $AA$ van der Waals couplings will affect the stability of our fractal spin liquid. Let us first consider the dual species case where these contributions can be neglected. We see in Fig.~\ref{fig:fractaldistance} that the first correction, coupling $A$ to $B$, occurs at $r=2a$. From this, we find that the fracton phase persists over approximately $400$ atoms (see \hyperref[methods]{Methods}). However, by interspersing the time-evolution with four sublattice pulses, this number increases to $1300$ (see \hyperref[methods]{Methods}), which can be systematically improved upon using local addressing. The corresponding expectation value of the stabilizers is $0.997$, giving a fidelity $\sqrt[N]{\mathcal F} \gtrapprox 0.998$. This establishes a high-fidelity implementation of a fracton state. Moreover, we note that the aforementioned procedure has the added benefit of cancelling nearest-neighbor $AA$ couplings, meaning this could also be successfully realized in single-species arrays if one can measure the $A$ sublattice.

In the Supplemental Materials \cite{SM}, we discuss a different lattice where measurements lead to an approximate realization of the paradigmatic X-cube model \cite{Vijay16} which can be realized with stabilizer expectation values $\approx 0.94$ in the absence of local addressing.

\begin{figure}
    \centering
    \begin{tikzpicture}
    \node at (0,0) {\includegraphics[scale=0.6]{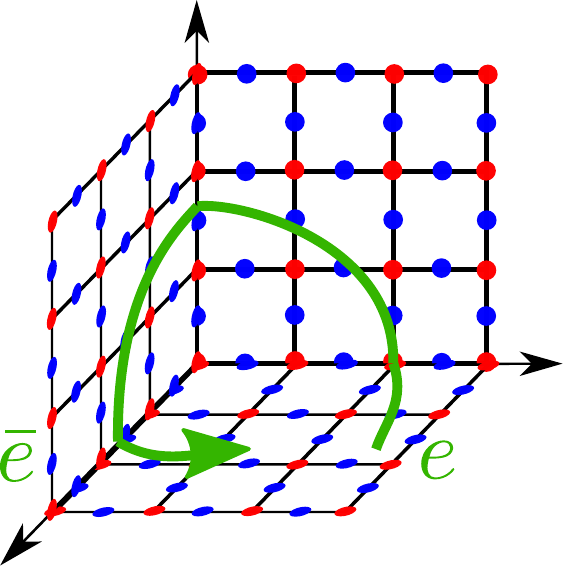}};
    \node at (4,0) {\includegraphics[trim={2cm 0 1cm 1cm},clip,scale=0.7]{2DsquareS3.pdf}};
    \node at (-1.5,1.35) {(a)};
    \node at (2.2,1.35) {(b)};
    \end{tikzpicture}
    \caption{\textbf{Towards $\mathbb Z_3$ and $S_3$ non-Abelian topological order.} (a) Time-evolving qutrits (formed by pairs of Rydberg-blockaded atoms) on the vertices and bonds of this 3D configuration and measuring the $A$ sublattice creates the $\mathbb Z_3$ toric code with a non-Abelian defect in the corner. This effectively realizes a square lattice with a disclination; an $e$-anyon traveling around it transforms into its conjugate $\bar e$. (b) Starting with the $\mathbb Z_3$ toric code on the $B$ sublattice, we load Rydberg atoms on the $C$ and $D$ sublattices. An appropriate sequence of time-evolution and pulses (see main text), followed by measuring the $C$ sublattice, produces non-Abelian $S_3$ topological order.}
    \label{fig:Z3_and_S3}
\end{figure}

\section{Non-Abelian defects and anyons \label{sec:nonabelian}}

The long-range entangled states we have obtained thus far are all Abelian, i.e., braiding their excitations can at most lead to phase factors. In this last section, we show how non-Abelian topological order can arise by using multiple time-evolution and measurement steps.

The first type of non-Abelian topological order we discuss is associated to the group $D_4$ (i.e., the symmetry group of the square). We show how this can be obtained by combining the toric and color code constructions discussed in Section~\ref{sec:toric}. Let us consider the honeycomb lattice in Fig.~\ref{fig:TC}(b), with the $A$ sublattice (red) at the center of plaquettes, the $B$ sublattice (blue) at the vertices, and the $C$ sublattice (purple) on the bonds. We find a strikingly simple representation of a state on the $C$ sublattice with non-Abelian $D_4$ topological order through the following sequence of controlled-$Z$ and single-site rotations (see \hyperref[methods]{Methods}):
\begin{equation}
\langle +|_{AB}\prod_{\langle b,c\rangle}CZ_{b,c} \; \prod_{b \in B} e^{-i \frac{\pi}{8} Y_b} \; \prod_{\langle a,b\rangle} CZ_{a,b} |+\rangle_{ABC}, \label{eq:D4}
\end{equation}
where $\langle b,c\rangle$ denotes nearest neighbors connecting the $B$ and $C$ sublattices. While Eq.~\eqref{eq:D4} is written in terms of post-selection for simplicity, we will explain that this is not required since measurement outcomes can be corrected. To create this state in Rydberg atom arrays, we first load the $A$ and $B$ sublattice into a product state, with controlled-$Z$ being naturally generated by time-evolution. As discussed in Section~\ref{sec:toric}, measuring the $A$ sublattice produces the color code, whose measurement-induced Abelian anyons can be removed with single-site rotations. We now apply the $Y$-rotation in Eq.~\eqref{eq:D4} and load the $C$ sublattice such that time-evolution induces the second controlled-$Z$. Finally, measuring the $B$ sublattice can again produce Abelian anyons, removable by single-site $X$-rotations.

Interestingly, if one had measured the $A$ sublattice at the end (or if one had not removed the anyons in the intermediate color code), a density of non-Abelian anyons could have been produced in the resulting $D_4$ topological order, which are much more difficult to efficiently remove. In this regard, the multi-step process is key, even when implementing Eq.~\eqref{eq:D4} in a quantum computer rather than through time-evolution in a Rydberg atom array.

We see that we can thus very efficiently obtain $D_4$ topological order. Although such a non-Abelian state is powerful, it does not allow for \emph{universal} fault-tolerant quantum computation. We now describe how to efficiently prepare a state with this property, namely $S_3$ non-Abelian topological order \cite{Mochon04}. While a measurement-based scheme has been discussed before \cite{Aguado08,Brennen09,Brell15}, here we present a scheme which is finite-depth (i.e., its duration independent of system size), and is moreover implementable in Rydberg atom arrays.

Our intermediate state will now be the $\mathbb Z_3$ toric code \cite{Kitaev_2003}, rather than the color code. For this, we need to work with qutrits instead of qubits. One straightforward option is to consider pairs of qubits where one uses only three states, which we label $|0\rangle,|1\rangle,|2\rangle$ (i.e., never initializing or evolving into the fourth state). In the context of Rydberg atoms, this is naturally achieved by having two atoms in close proximity such that the Rydberg blockade \cite{Jaksch00,Lukin01} forbids them from being simultaneously excited (although using multiple energy levels in a single atom offers an alternative route to qutrit encoding). Similar to the qubit case, time-evolution can generate the controlled-$Z$ gate for qutrits, defined as $CZ|i \rangle \otimes |j\rangle = \omega^{ij} |i \rangle \otimes |j\rangle$ with $\omega = e^{\frac{2\pi i}{3}}$ (see \hyperref[methods]{Methods}). Hence, placing our qutrits on the lattice in Fig.~\ref{fig:TC}(a) can generate the $\mathbb Z_3$ toric code on the honeycomb lattice. However, for our purposes, it will be convenient to instead consider the Lieb lattice (i.e., vertices ($A$) and bonds ($B$) of the square lattice), thereby producing this topological state on the $B$ sublattice.

As an aside, we note that although this topological order is Abelian, it can be used to trap non-Abelian defects. For instance, consider the 3D configuration in Fig.~\ref{fig:Z3_and_S3}(a), which can be interpreted as a 2D Lieb lattice with a disclination. Repeating the above preparation protocol produces the $\mathbb Z_3$ toric code with such a disclination defect. If an $e$- or $m$-anyon moves around the corner in Fig.~\ref{fig:Z3_and_S3}(a), it will transform into the conjugate anyon ($\bar e$ or $\bar m$), leading to a quantum dimension $d=3$ for the defect (see \hyperref[methods]{Methods}).

Finally, to transform this state into $S_3$ topological order, we consider the $C$ and $D$ sublattices defined in Fig.~\ref{fig:Z3_and_S3}(b), which we load with effective qubits. Through time-evolution and sublattice rotations, we can first implement controlled-charge-conjugation for the nearest neighbors connecting the $B$ and $C$ sublattices (see \hyperref[methods]{Methods}). Subsequently we implement controlled-$Z$ for nearest neighbors connecting the $C$ and $D$ sublattices. Remarkably, measuring the $C$ qubits produces the ground state of the $S_3$ quantum double model \cite{Kitaev_2003} on the remaining qutrits on $B$ and qubits on $D$ \cite{SM}.

At a conceptual level, this construction can be interpreted as effectively gauging the $\mathbb Z_2$ charge-conjugation symmetry of the $\mathbb Z_3$ toric code, in line with $S_3 = \mathbb Z_3 \rtimes \mathbb Z_2$. In a companion work, we explore more generally how gauging can be implemented using finite-depth time-evolution and measurements \cite{framework}.

\section{Outlook}
In summary we have shown how a variety of highly sought after long-range entangled states can be efficiently created from simple ingredients: initialization of qubits, unitary time-evolution under Ising interactions, and measurement of a subset of qubits, which can be realized on existing experimental platforms. We have outlined a blueprint for implementation in Rydberg atom arrays, by identifying lattice structures that leverage their intrinsic interactions and demonstrate high fidelity preparation of 1D GHZ and 2D toric and color code states, as well as 3D fracton phases. The last---which has been discussed extensively in theory---now seem within experimental reach. We have also discussed several possible generalizations of our protocol. For instance, using the Rydberg blockade one can realize long-range entangled {\em qutrits}; using 3D configurations one can obtain 2D lattices with exotic defects; and by appealing to multiple measurement steps, one can even realize states hosting non-Abelian anyons.

Given that our protocol is within experimental reach, it is a worthwhile endeavour to further explore how the resulting states could be used. The 1D GHZ state is well-known to be useful in quantum metrology due to its nonclassical sensitivity to external fields. We have outlined how one can create surface codes for the toric and color codes, even allowing for active error correction. Similarly, our proposed 3D fracton states could be similarly used to perform quantum computation, although its surface code properties have not yet been extensively explored---perhaps our concrete proposal and its imminent implementation can be a catalyst for such further work.

A particularly exciting prospect is the realization of non-Abelian topological order in engineered quantum systems. Here we have demonstrated how both $S_3$ and $D_4$ topological order can be realized in Rydberg atom arrays. It is worth noting that our novel schemes are more broadly applicable. E.g., our proposal could be used to realize $D_4$ topological order in a chip of superconducting qubits using the simple geometry of Fig.~\ref{fig:TC}(b). Note that Eq.~\eqref{eq:D4} requires a circuit which is only ten layers deep, which is achievable with current coherence times. We thus conclude that various exotic long-range entangled states discussed in the present work can be prepared in multiple existing platforms, providing a thrilling view of what is to come.

\vspace{5pt}


\begin{acknowledgements}
RV thanks Hannes Bernien for an inspiring discussion about dual-species Rydberg arrays. NT is grateful to Wenjie Ji and Sagar Vijay for helpful discussions on quantum double models and gauging. Moreover, the authors thank Ryan Thorngren for collaboration on the companion work Ref.~\cite{framework}. DMRG simulations were performed using the TeNPy Library \cite{Hauschild18}. RV is supported by the Harvard Quantum Initiative Postdoctoral Fellowship in Science and Engineering. NT is supported by NSERC. AV and RV are supported by the Simons Collaboration on Ultra-Quantum Matter, which is a grant from the Simons Foundation (618615, A.V.).
\end{acknowledgements}

\bibliography{main.bbl}

\small

\section*{Methods \label{methods}}

Written out in spin notation, the Hamiltonian of interest is:
\begin{equation}
H =  \sum_{\bm i} \frac{\Omega_{\bm i}}{2} X_{\bm i} - \sum_{\bm i} \frac{h_{\bm i}}{2} Z_{\bm i} + \frac{1}{8} \sum_{\bm i \neq \bm j} V_{ij} Z_{\bm i} Z_{\bm j} \label{eq:H}
\end{equation}
with $h_{\bm i} = \delta_{\bm i} + \frac{1}{2} \sum_{\bm{ j} \neq \bm i} V_{ij}$ (compared to Eq.~\eqref{eq:Hryd}). We note that this extra shift of the longitudinal field---explicitly giving us an Ising interaction---is also practically significant, since it is key in some of the symmetry arguments which we used to argue stability of the resulting state. (We note that various theoretical works \cite{Fendley04,Samajdar20square,PhysRevX.10.021057,VerresenLukinVishwanath21,Samajdar2021quantum,Myerson21,Slagle21Rydberg} have already studied the ground state physics of Rydberg atom arrays described by Eq.~\eqref{eq:H}; here we do not consider ground states, but rather states obtained by time-evolution.)

\begin{center}\textbf{Dual species and F\"orster reasonances}\end{center}

One interesting option is to use different Alkali atoms for the two sublattices of our set-ups \cite{Beterov15,Samboy17,Otto20}. In addition to making it possible to measure a single sublattice, it can be used to suppress the intra-species interaction. In particular, it is known that by using (near-)F\"orster resonances, one can choose the Rydberg levels of each atom such that the intra-species interaction is negligible. For instance, the inter-species interaction between the $48S_{1/2}$ state in Rubidium and the $51S_{1/2}$ state in Cesium is roughly 100 times larger than the intra-species coupling \cite{Beterov15}, which could potentially be further optimized by exploring higher Rydberg states or other atoms.

Here a technical note is in order.
The tuneability of the dual species interaction derives from finding Rydberg states such that there are nearly-degenerate states (e.g., for a given choice of $S$-states in Rubidium and Cesium, the energy might be approximately unchanged if the Rubidium (Cesium) atom in an $S$-state transitions to a lower (higher) energy $P$-state).
For our purposes, it is desirable to place the atoms far apart enough such that this transition is considerably off-resonant; beyond the so-called `crossover disance' $R_c$, the interaction is effectively van der Waals (for distances shorter than $R_c$, the interaction decays as $\sim 1/r^3$ and we would have additional spin-flop terms in the Hamiltonian, which we want to avoid) \cite{Beterov15}. For instance, for the aforementioned example of the Rb $48S_{1/2}$ and Cs $51S_{1/2}$ states, $R_c \approx 7.5 \mu \textrm{m}$. For a spacing $a= 10 \mu \textrm{m}$, we have that the inter-species interaction is still at the MHz scale (with the intra-species coupling being negligible). As we will see, our preparation time is $\frac{\pi}{V(a)}$ which is thus on the order of microseconds---fast enough to neglect decoherence effects which arise at $\sim 100\mu \textrm{s}$. The preparation time can be reduced by considering higher principal quantum numbers.

\subsection*{Lower bound on stabilizer fidelity pre-measurement}

Ref.~\cite{Cramer10} proved that the a quantum state $|\psi\rangle$ on $N$ qubits has a fidelity $\mathcal F \geq 1-N\varepsilon/2$ for a stabilizer state corresponding to stabilizer operators $K$ if $\langle \psi | K |\psi\rangle = 1-\varepsilon$. If $N\varepsilon \ll 1$, this implies that the fidelity density satisfies the lower bound $\sqrt[N]{\mathcal F} \geq 1-\varepsilon/2$. Since the fidelity density will converge a finite value as $N \to \infty$ (at least for a state with a finite correlation length), we expect that the condition $N \varepsilon \ll 1$ does not need to be explicitly enforced (although this is of course the region where the bound is the most useful). Indeed, in one spatial dimension one can straightforwardly prove using matrix product state techniques that the bound $\sqrt[N]{\mathcal F} \geq 1-\varepsilon/2$ is \emph{always} satisfied for $\langle \psi | K |\psi\rangle = 1-\varepsilon$, even in the thermodynamic limit.

\subsection*{Lower bound on stabilizer fidelity post-measurement}

The above gives us a lower bound on the fidelity for the cluster state, based on the expectation value of the stabilizer. We are ultimately interested in the fidelity with respect to the long-range entangled state for the post-measurement state. Fortunately, these two can be equated for the following reason. Firstly, in all our examples, one can show that $\langle X_n \rangle =0$, making both measurement outcomes equally likely. Secondly, we find that the correlations between $X$-measurements on different sites is completely negligible (e.g., in the 1D case we find that $\langle X_n X_{n+2} \rangle < 10^{-10}$ even upon including all longer-range couplings in the single-species set-up). Hence, to a very good degree, the post-measurement state is given by:
\begin{equation}
|\psi_\textrm{out}\rangle = \prod_{a \in A} \left( \frac{X_a + s_a}{\sqrt{2}} \right) |\psi\rangle,
\end{equation}
where $s_a = \pm 1$ denotes the measurement outcome. From the above explicit formula, it is straightforward to confirm our claim about the equality of fidelities.

\subsection*{Numerical simulation of cluster state preparation}

We consider $^{87}$Rb in the Rydberg level $70S_{1/2}$. The intrinsic van der Waals interaction is $H_\textrm{int} = U\sum_{i<j} n_i n_j/|r_i-r_j|^6$ with $U \approx 5 \textrm{THz} \; \mu \textrm{m}^6$. If $a$ is the lattice spacing, then the nearest-neighbor interaction is $V(a) = U/a^6$. For $a = 10 \mu\textrm{m}$ ($a =12 \mu\textrm{m}$) we have $V(a) \approx 5\textrm{MHz}$ ($V(a) \approx 1.7 \textrm{MHz}$). We initialize all atoms in their ground state. We then apply an $X$-pulse for a time $t = t_\textrm{pulse}$ with strength $\Omega = \frac{\pi}{2t_\textrm{pulse}}$. We take $t_\textrm{pulse} = 20 \textrm{ns}$, corresponding to $\Omega \approx 78 \textrm{MHz}$. We then apply a $Z$-pulse for the same time with $h = -\frac{\pi}{2 t_\textrm{pulse}}$. (Note that one can avoid the need for this separate pulse: since $Z$ commutes with the Rydberg interaction, one can perform both at the same time.) After these pulses, we (roughly) realize $|-\rangle^{\otimes N}$ (up to corrections due to the interactions which are present during the short pulse; our simulations include these interactions). We then time-evolve under the natural Ising interaction ($\Omega=h=0$) for a time $t_\textrm{SPT} = \frac{\pi}{V(a)} - \frac{5}{2} t_\textrm{pulse}$. 
After this time-evolution, we apply a final $X$-pulse of strength $\Omega = - \frac{\pi}{2 t_\textrm{pulse}} \times \frac{1}{2^6}$. 
The simulations are performed using the density matrix renormalization group method for infinite systems \cite{White92}, thereby directly working in the thermodynamic limit. We found that small bond dimensions, $\chi = 20$ -- $40$, were already sufficient to guarantee convergence. (Note: if we could reduce the pulse time to e.g., $t_\textrm{pulse}=5 \textrm{ns}$, then the fidelity goes up another order of magnitude: we find $0.999997$ (for $a=12\mu \textrm{m}$), in close agreement with the prediction in Eq.~\eqref{eq:ideal}).

\subsection*{Dual stabilizer}

When measuring $X_b$ on a given site of the $B$ sublattice, then any stabilizer that does not commute with it is no longer a stabilizer of the post-measurement state. However, sometimes a product of such stabilizers commutes with all $X$-measurements, thereby providing a constraint on the post-measurement state. E.g., in the 1D case, the product of $X_{2n} Z_{2n+1} X_{2n+2}$ (each individual factor not commuting with the $X$-measurements) gives us the constraint that the resulting post-measurement state has the $\mathbb Z_2$ symmetry $\prod_n Z_{2n+1}$. In the 2D toric code, we instead get a local constraint: the product of the cluster term on four bonds around a plaquette give us the condition $\mathcal B_p = +1$.

\subsection*{The effect of symmetry-preserving couplings}

For the case of the GHZ and toric code states, we claimed that $AA$ and $BB$ Ising couplings do not affect the resulting phase of matter. The argument for $BB$ is simple to see: these gates commute with the measurement and hence dress the post-measurement stat with a finite-depth circuit (which cannot destroy the long-range entangled states). The $AA$ couplings are more interesting and subtle: in fact, they also push through to just being effective finite-depth circuits on the post-measurement states, which one can argue by using the stabilizer property of the cluster state, which allows one to replace it by $X$ (or a product thereof) on the $B$ sublattice. This is derived in explicit detail for the 1D case in the Supplemental Material \cite{SM}. In conclusion, only the longer-range $AB$ couplings affect the quality of the resulting phase of matter.

\subsection*{Improvements using local addressing}

Our protocol can be further improved using spatially-dependent fields, called local addressing \cite{Labuhn14}. This tool has already been used in large Rydberg arrays \cite{Omran19}.

To illustrate how this can suppress or even cancel unwanted van der Waals interactions, let us return to the 1D case (Section~\ref{sec:1D}). Using spatially-dependent $X$-pulses, we can flip qubits on a sublattice of our choosing. Let $\mathcal X_k$ correspond to flipping every fourth site starting from $k=1,2,3,4$. If $U(t)$ denotes the time-evolution, we consider the following alternation of time-evolution and sublattice pulses sketched in Fig.~\ref{fig:1D}(d):
\begin{equation}
\mathcal X_2 \mathcal X_3\;
U(t_\textrm{SPT}/2) \;
\mathcal X_1 \mathcal X_3 \;
U(t_\textrm{SPT}/2) \;
\mathcal X_1 \mathcal X_2 \;
U(t_\textrm{SPT}).
\end{equation}
The additional procedure does not affect the net time-evolution for any sites separated by an odd multiple of lattice spacings; in particular, the nearest-neighbor evolution still prepares the cluster state. However, the second-nearest neighbor coupling drops out completely, making $r=3a$ the dominant correction. (We note that the fourth-nearest neighbor experiences an effective \emph{doubled} time-evolution.) Similarly, local addressing can suppress unwanted contributions for the fracton model (see next header).

We note that this cancellation procedure can also be useful \emph{after} the state preparation. Indeed, the remaining $B$ sublattice will continue to time-evolve under the long-range Ising interaction. By construction, this coupling is small, especially in the dual-species set-up. Nevertheless, such effects can accumulate. One can then apply local pulses as above to effectively change the sign of a problematic coupling, such that further time-evolution will undo its effects. In the 1D case this is not necessary, since the GHZ state is an eigenstate of the van der Waals interactions.

\subsection*{Fracton order}

Upon measuring the red sublattice of the cluster state on the 3D lattice in Fig.~\ref{fig:fractaldistance}(a), we obtain the following stabilizers:
\begin{center}
\includegraphics[scale=0.62]{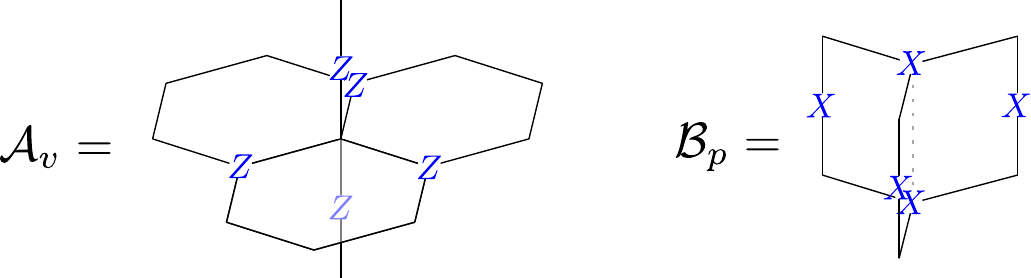}
\end{center}
These can be equated with the stabilizers of the fractal spin liquid introduced in Ref.~\cite{Yoshida13}.

We obtain the following expectation value for the stabilizers in the dual-species set-up:
\begin{align}
\langle \mathcal A_v \rangle &\approx \cos^{15}\left(  \frac{\pi}{2 \cdot 2^6} \right) 
\cos^{6}\left(  \frac{\pi}{2 \cdot \sqrt{5}^6} \right) \cdots  \approx 0.995\\
\langle \mathcal B_p\rangle &\approx \cos^{42}\left(  \frac{\pi}{2 \cdot 2^6} \right)  \cos^{12}\left(  \frac{\pi}{2 \cdot \sqrt{5}^6}  \right)  \cdots \approx 0.986
\end{align}

We can improve this by using local addressing: we can cancel the out-of-plane contributions at $r=2a$ by noticing that the red sublattice is tripartite (forming a triangular lattice in each layer). Correspondingly, let us split $A$ up into $A_1 \cup A_2 \cup A_3$, and let $\mathcal X_{A_i}$ denote flipping qubits on the $A_i$ sublattice via an $X$-pulse. Consider
\begin{equation}
\mathcal X_{A_3} U(t)\mathcal X_{A_3} \mathcal X_{A_2} U(t)\mathcal X_{A_2}
\mathcal X_{A_1} U(t)\mathcal X_{A_1}
U(t)
\end{equation}
Setting $t=t_\textrm{SPT}/2$ will create the cluster state, and we have effectively removed the aforementioned couplings. We now obtain the following improved results:
\begin{align}
\langle \mathcal A_v \rangle &\approx \cos^{3}\left(  \frac{\pi}{2 \cdot 2^6} \right) 
\cos^{6}\left(  \frac{\pi}{2 \cdot \sqrt{5}^6} \right) \cdots  \approx 0.9985\\
\langle \mathcal B_p\rangle &\approx \cos^{6}\left(  \frac{\pi}{2 \cdot 2^6} \right)    \cos^{12}\left(  \frac{\pi}{2 \cdot \sqrt{5}^6}  \right) \cdots  \approx 0.997
\end{align}

\subsection*{$D_4$ topological order}

Here we prove that Eq.~\eqref{eq:D4} exhibits $D_4$ non-Abelian topological order. First, we consider
\begin{equation}
\langle +|_A \prod_{\langle a,b\rangle } CZ_{a,b}|+\rangle_{A,B}.
\end{equation}
As discussed in Section~\ref{sec:toric} of the main text, this is the color code on the $B$ sublattice. Next, we want to use the fact that gauging the $\mathbb Z_2$ Hadamard symmetry of the color code produces $D_4$ topological order \cite{PremWilliamson2019}. Usually, gauging is a nonlocal transformation which would require a linear-depth circuit. However, in our companion work \cite{framework}, we show that an arbitrary initial state can be gauged by a finite-depth circuit (made of controlled-Z gates) and single-site projections or measurements. Here we use this general fact to produce $D_4$ topological order. In particular, we first perform a $Y$ basis rotation to transform the Hadamard symmetry into $\prod_{b \in b} X_b$, after which we entangle the $B$ degrees of freedom with a product state on the $C$ sublattice through controlled-$Z$:
\begin{equation}
\prod_{\langle b,c \rangle} CZ_{b,c} \prod_{b\in B} e^{-i \frac{\pi}{8} Y_b}\langle +|_A \prod_{\langle a,b\rangle } CZ_{a,b}|+\rangle_{A,B}.
\end{equation}
Now projecting the $B$ sites into $\langle +|$ will effectively implement the gauging of the original color code, thereby producing $D_4$ topological order.

\subsection*{$\mathbb Z_3$ toric code, its defects, and $S_3$ topological order}

To create the $\mathbb Z_3$ toric code \cite{Kitaev_2003}, we place the qutrits on the vertices and bonds of the square lattice, forming the $A$ and $B$ sublattices respectively. One can show that combining the natural Rydberg interaction with several sublattice pulses gives the following effective nearest-neighbor interaction \cite{SM}:
\begin{equation}
H_{a,b} = U \left(  n_{1,a} n_{1,b} + n_{2,a} n_{2,b} -n_{1,a} n_{2,b} - n_{2,a} n_{1,b}  \right), \label{eq:1}
\end{equation}
where $n_{i,\lambda}$ denotes whether atom $i$ on site $\lambda$ is excited. Hence, the time-evolution $e^{- \frac{4\pi i}{3 U} H_{a,b}}$ implements the controlled-$Z$ gate defined in the main text, producing the topological state. Note that the first correction connecting different sublattices has a prefactor $\left(\frac{r_{AB}}{r_{AB'}}\right)^6 = \frac{1}{\sqrt{5}^6} \approx 0.008$, justifying a nearest-neighbor approximation.

If our lattice has a disclination, it can trap a non-Abelian defect. Indeed, the usual $\mathbb Z_3$ cluster state is prepared by having a pattern of $CZ$ and $(CZ)^\dagger$ \cite{Brell15}
being unitarily equivalent to using only $CZ$ if we act with charge-conjugation on, for example, every other $A$ site along with the $B$ sites immediately above and to the right. However, a disclination destroys the bipartiteness of $A$, leading to a charge-conjugation defect. This kind of defect has a quantum dimension $d=3$ \cite{BarkeshliBondersonChengWang2019}.

In the Supplemental Materials we describe in detail how this $\mathbb Z_3$ toric code can be transformed into non-Abelian topological order. We briefly summarize the key steps here. First, starting with the $\mathbb Z_3$ toric code on the $B$ sublattice in Fig.~\ref{fig:Z3_and_S3}(b), we load the $C$ sublattice with Rydberg atoms serving as qubits. Their natural interaction is:
\begin{equation}
H_{BC} = U_{BC} \sum_{\langle b,c\rangle} \left( n_{1,b} n_c + n_{2,b} n_c \right). \label{eq:2}
\end{equation}
Combining time-evolution for a time $t = \frac{\pi}{U_{BC}}$ with an appropriate pulse effectively implements controlled-charge-conjugation for nearest-neighbors. We then load the $D$ sublattice with qubits, and let $C$ and $D$ interact under
\begin{equation}
H_{CD} = U_{CD} \sum_{\langle c,d \rangle} n_{c} n_{d} \label{eq:3}
\end{equation}
for time $t = \frac{\pi}{U_{CD}}$, combined with a pulse that cancels out coupling to $B$ sublattice, implementing a controlled-$Z$. Finally, measuring the $C$ qubits produces the ground state of the $S_3$ quantum double model \cite{Kitaev_2003} on the remaining qutrits on $B$ and qubits on $D$ \cite{SM}.


\onecolumngrid
\appendix
\setcounter{figure}{0}
\renewcommand\thefigure{\thesection.\arabic{figure}}

\section{Preparation of cat state with timing imperfections}

Consider the state $|\psi(t)\rangle = e^{-iHt}|-\rangle^{\otimes N}$ obtained by evolving the product state with the Ising Hamiltonian $H = J \sum Z_n Z_{n+1}$. (For the special time $t_\textrm{SPT} = \frac{\pi}{4 |J|}$, this coincides with the cluster state.) 

\begin{shaded}
\textbf{Theorem.} Suppose we measure the even sites of $|\psi(t)\rangle$ in the Pauli-$X$ basis, producing the state $|\psi_\textrm{out}(t)\rangle$ for a given measurement outcome. Then the remaining odd sites have a correlation function
\begin{equation}
\langle \psi_\textrm{out}(t)| Z_{2m-1} Z_{2(m+n)-1} |\psi_\textrm{out}(t) \rangle \sim e^{-n/\xi}. \label{eq:postmeasurement_xi}
\end{equation}
Define $\alpha = \cos^2(2t)$. If $n\gg \frac{1-\alpha}{1+\alpha}$, then the correlation length $\xi$ is normally distributed (corresponding to the randomness of measurement), $\xi \sim N(\bar \xi, \sigma_\xi)$, with the following mean and standard deviation:
\begin{equation}
\bar \xi = \frac{2}{(1+\alpha)\ln\left( \frac{1+\alpha}{1-\alpha} \right)} \qquad \textrm{and} \qquad \sigma_\xi = \frac{\bar \xi}{\sqrt{n}} \sqrt{ \frac{1-\alpha}{1+\alpha}}. \label{eq:normal}
\end{equation} 
\end{shaded}

\begin{figure}
    \centering
    \includegraphics[scale=0.33]{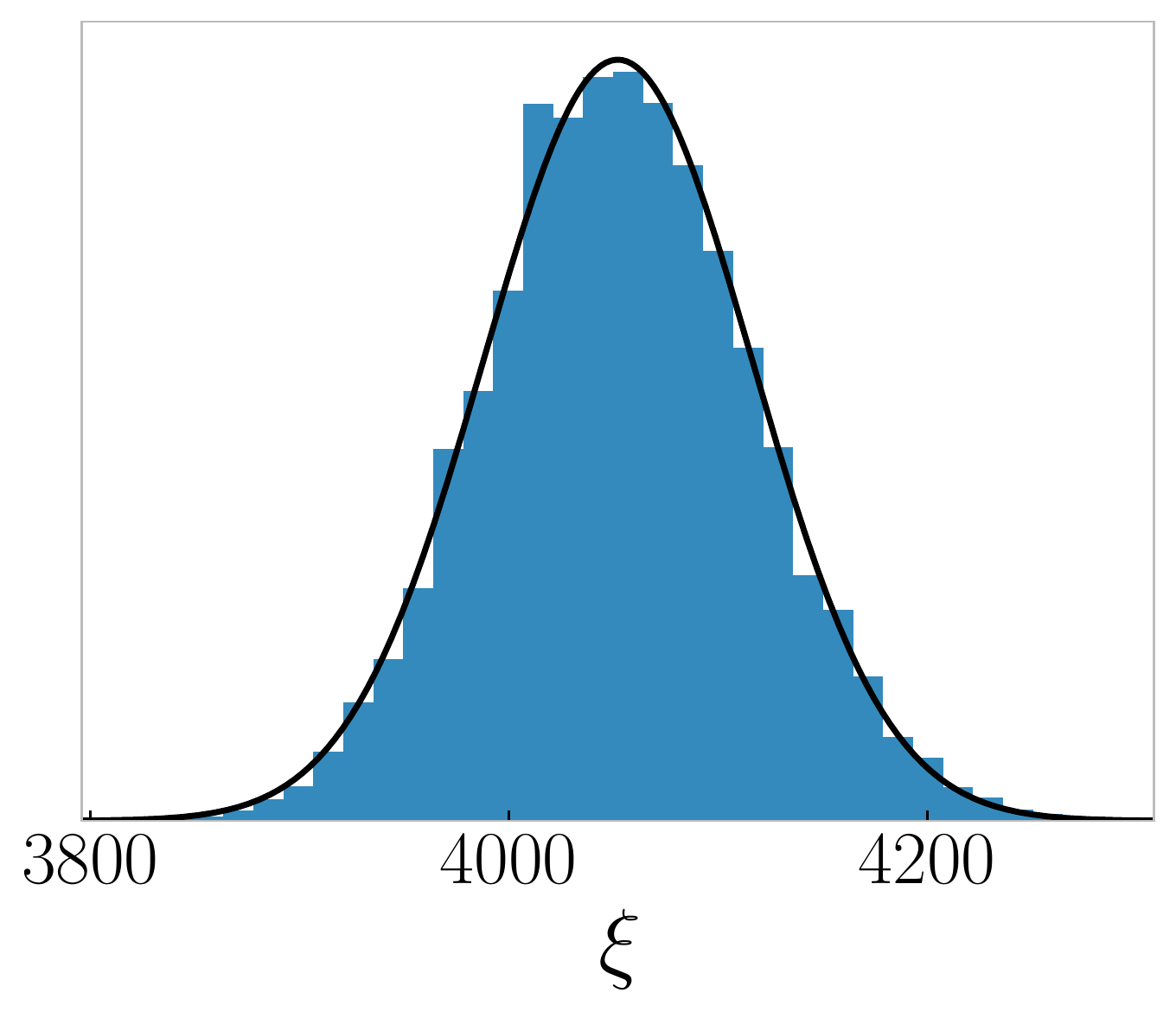}
    \caption{The distribution of the correlation length $\xi$ over a distance $n = \bar \xi$ for $t=0.99 \times t_\textrm{SPT}$ (see Eq.~\eqref{eq:postmeasurement_xi}. The black line is the predicted normal distribution (see Eq.~\eqref{eq:normal}).}
    \label{fig:distribution_1pct}
\end{figure}

Before proving this result, let us make a few observations:
\begin{enumerate}
    \item For the special case $t=0$ (i.e., $\alpha = 1$), we recover $\bar \xi =0$ for the product state, whereas the cluster state at $t=t_\textrm{SPT}$ (i.e., $\alpha = 0$) gives rise to a true cat state with $\bar \xi = \infty$.
    \item Expanding around $t = t_\textrm{SPT} + \delta t$ (i.e., around $\alpha = 0$), we find $\xi = \frac{1}{\alpha} - 1 + O(\alpha)$. The correlation length thus blows up as $\xi \sim \frac{4 t_\textrm{SPT}^2}{\pi^2} \frac{1}{\left(\delta t\right)^2}$ upon approaching $t \to t_\textrm{SPT}$.
    \item For all $t$, we see that $\sigma_\xi \to 0$ as $n \to \infty$. However, for practical purposes it can be meaningful to set $n = \bar \xi$ to get a sense of how the correlation length will vary over physically relevant length scales. An example is shown in Fig.~\ref{fig:distribution_1pct} for $t = 0.99 t_\textrm{SPT}$, where we find that the deviations around $\bar \xi \approx 4000$ are small.
    \item The plot in Fig.~\ref{fig:1D}(b) in main text corresponds to $\bar \xi$ with a thickness of one standard deviation $\sigma_\xi$, where we have chosen $n  =\bar \xi$.
\end{enumerate}

\begin{proof}
Firstly, since we are time-evolving with a purely-diagonal nearest-neighbor Ising Hamiltonian, the $X$-correlations are uncorrelated beyond two sites. More precisely, $\langle \psi(t)|\prod_{n \in \mathcal S} X_{2n}  |\psi(t)\rangle =\prod_{n \in \mathcal S} \langle \psi(t)|  X_{2n} |\psi(t)\rangle$ for any set of operators that does not include \emph{all} even operators. Hence, if we measure $X$ on all even sites (leaving out one for convenience), then the state with outcome $X_{2n} = s_{2n}$ can be written as
\begin{equation}
|\psi_\textrm{out}(t) \rangle = \prod_n \frac{X_{2n} + s_{2n}}{\sqrt{2-2\alpha s_{2n} }} |\psi (t)\rangle \qquad \textrm{where } \alpha := |\langle X \rangle| = \cos^2(2t).
\end{equation}
From this, one can derive that
\begin{equation}
|\langle \psi_\textrm{out}| Z_{2m-1} Z_{2n+1} |\psi_\textrm{out} \rangle| = \prod_{k=m}^n \left( \frac{1-\alpha}{1+\alpha}\right)^{\frac{1-s_{2k}}{2}}.
\end{equation}
For instance, this follows from the factorization property $\langle Z_{2m-1} Z_{k} Z_{k} Z_{2n+1} \rangle =\langle Z_{2m-1} Z_{k}\rangle \langle Z_{k} Z_{2n+1} \rangle  $ and a direct calculation for the simplest case $n=m$.

Hence, we see that the correlation length on the odd sites is given by
\begin{equation}
\xi = \frac{1}{\ln\left( \frac{1+\alpha}{1-\alpha} \right)} \times \frac{1}{\rho}
\end{equation}
where $\rho$ is the fraction of measured sites between the two $Z$-operators which have an outcome $s = -1$. If we consider all possible measurement outcomes, then $\rho \times n$ is described by a binomial distribution $B(n,p)$ where $n$ is the number of even sites between two $Z$-operators, and $p = \frac{1+\alpha}{2}$ is the probability of having outcome $s= -1$. Hence, for large $n$, $\rho$ is described by the normal distribution $N(p,\sigma)$ with standard deviation $\sigma = \sqrt{ \frac{p(1-p)}{n}}$. Hence, if $\sigma \ll p$ (i.e. $n \gg \frac{1}{p}-1 = \frac{1-\alpha}{1+\alpha}$), then $1/\rho$ is normally distributed as $N(1/p,\sigma/p^2)$. QED
\end{proof}

\section{Details for preparing cluster and GHZ state in 1D \label{app:GHZ}}

Consider 1D chain with $N$ atoms with spacing $a$:
\begin{equation}
\boxed{ H(\Omega,h) = \frac{\Omega}{2} \sum_{n=1}^N X_{n} - \frac{h}{2} \sum_{n=1}^N Z_{n} + \frac{V(a)}{4} \sum_{n=1}^N \sum_{k \geq 1} v_k Z_n Z_{n+k} \quad \textrm{with } 
h = \delta + V(a) \sum_{k \geq 1} v_k } \label{eq:H}
\end{equation}
In our nearest-neighbor model, we have $v_{n>1}=0$; in the single-species model we have $v_n = 1/n^6$; in the dual species (where even and odd sites have different species), we have $v_{2n-1} = 1/(2n-1)^6$ and $v_{2n} = 0$. However, our formulas will be valid for general $v_k$. 

We will be interested in time-evolving with the interaction term over a time $t_\textrm{SPT} = \frac{\pi}{V(a)}$. In particular, we consider the following unitary operator:
\begin{equation}
U_\textrm{SPT} := e^{-i t_\textrm{SPT} H(0,0)} = e^{- \frac{i \pi}{4} \sum_n \sum_{k\geq 1} v_k Z_n Z_{n+k} }
\end{equation}
We will refer to this as the SPT-pulse. Indeed, in the nearest-neighbor model $v_{k} = \delta_{k,1}$, this would create the ideal cluster SPT phase, a state which we denote by $|\textrm{cluster}\rangle$ and which is the ground state of the cluster model, $H_\textrm{cluster} = -\sum_n Z_{n-1} X_n Z_{n+1}$.

\subsection{Realizing the cluster state}

Starting from the ground state of $H(+\infty,0)$ and then applying the SPT-pulse, the state we end up creating is:
\begin{equation}
|\psi\rangle = U_\textrm{SPT} |-\rangle^{\otimes N} = \prod_n \prod_{k\geq 2}\left[ \cos\left( \frac{\pi v_k}{4} \right)  - i \sin\left( \frac{\pi v_k}{4} \right) Z_n Z_{n+k} \right] |\textrm{cluster}\rangle \label{eq:SPT}
\end{equation}
From this, we can derive various \emph{exact} properties of the resulting state:
\begin{align}
\mathcal F_\textrm{SPT} = \sqrt[N]{|\langle \textrm{cluster}|\psi\rangle|^2 } & \geq \prod_{k=2}^\infty \cos^2 \left( \frac{\pi v_k}{4} \right) \\
\langle \psi| Z_{m-1} X_{m} Z_{m+1} |\psi\rangle &= \prod_{k=2}^\infty \cos^2 \left( \frac{\pi v_k}{2} \right) \\
\langle \psi| X_n |\psi \rangle = \langle \psi| Y_n |\psi \rangle = \langle \psi| Z_n |\psi \rangle &= 0 \\
\langle \psi| X_{n} X_{n+2} |\psi \rangle &= \frac{c_2^2 t_3^2}{2}
\sum_{\sigma=0,1} (-1)^\sigma \prod_{k= 3}^\infty \left[ c_k^4 \left(1 + (-1)^\sigma t_{k-1} t_{k+1}\right)^2 \right] .
\end{align}
The inequality in the first formula is true for $v_k$ which decay fast enough with $k$ (such as in our Rydberg case $v_k = 1/k^6$). The other three formulas are true for all $v_k$.
In the latter formula, we introduced the shorthand $c_k := \cos(\pi v_k/2)$ and $t_k := \tan(\pi v_k/2)$.

These formulas are straightforwardly evaluated. Let us first consider the single-species model $v_k = 1/k^6$. In this case, we find a fidelity per site of $\mathcal F_\textrm{SPT} \gtrapprox 0.99985$ (actually turns out to be approximate equality), i.e., we are \emph{very close} to the ideal cluster state. E.g., even for $N=100$ atoms, we have the fidelity $|\langle \textrm{cluster}|\psi\rangle|^2 \approx 0.985$. Relatedly, the stabilizer defining the cluster state also has a very large value: $\langle \psi| Z_{m-1} X_n Z_{m+1} |\psi \rangle \approx 0.9994$. Moreover, we find that
\begin{equation}
\langle \psi| X_n |\psi \rangle = 0 \qquad \textrm{and} \qquad |\langle \psi| X_n X_{n+2} |\psi\rangle| < 10^{-10}. \label{eq:uncorrelated}
\end{equation}
This tells us that measuring in the $X$-basis on, say, the even sites essentially corresponds to a completely uncorrelated flat distribution (to an extremely good approximation).


\subsection{Creating the GHZ state through measuring half the system}

Let us now measure $X_n$ on every odd site. Due to Eq.~\eqref{eq:uncorrelated}, the resulting state is
\begin{equation}
|\psi_\textrm{proj}\rangle \approx \prod_n \frac{s_{2n-1}+X_{2n-1}}{\sqrt{2}} |\psi \rangle,
\end{equation}
where $s_n$ is the measurement outcome. Note that after judiciously chosen single-site flips, we can write
\begin{equation}
|\psi_\textrm{proj}\rangle \approx \prod_n \frac{1+X_{2n-1}}{\sqrt{2}} |\psi \rangle.
\end{equation}
From this, it follows that
$|\langle \textrm{GHZ}|\psi_\textrm{proj}\rangle|^2 \geq |\langle \textrm{SPT}|\psi\rangle|^2$. I.e., a good fidelity for the cluster state implies a good fidelity for the GHZ state after preparation! E.g., for the above cluster fidelity of $\approx 0.985$ for 100 atoms, we find a lower bound $|\langle \textrm{GHZ}|\psi_\textrm{proj}\rangle|^2 \geq 0.97$. However, it turns out that this underestimates the true result. In addition, this can be improved by noting that the correction due to the second-nearest neighbor can be removed by a single-site rotation, as we discusss now.

\subsection{Modified stabilizer}

We consider the effect of $v_2$, leading to the state
\begin{equation}
|\psi\rangle = \exp\left(-\frac{i\pi v_2}{4} \sum Z_n Z_{n+2}\right) |\textrm{cluster}\rangle.
\end{equation}

This is still a $\mathbb Z_2 \times \mathbb Z_2$ SPT, so measuring, say, the even sites should give a cat state on the odd sites. To see this, let us rewrite
\begin{align}
|\psi \rangle = \prod_n \exp\left(-\frac{i\pi v_2}{4} Z_{2n-1} Z_{2n+1}\right)  \prod_n \exp\left(-\frac{i\pi v_2}{4} X_{2n+1}\right)  |\textrm{cluster}\rangle
\end{align}

Note there are no operators acting on the even sites. Hence, if we measure on the even sites, the projectors that project the state onto our measurement outcome commute with the above operators. The result is thus
\begin{equation}
|\psi_\textrm{out}\rangle = \prod_n \exp\left(-\frac{i\pi v_2}{4} Z_{2n-1} Z_{2n+1}\right)  \prod_n \exp\left(-\frac{i\pi v_2}{4} X_{2n+1}\right)  |\textrm{cat}\rangle = U|\textrm{cat}\rangle,
\end{equation}
where the $|\textrm{cat}\rangle$ state is in the diagonal basis (its precise configuration depending on the measurement outcome). This is a demonstration of the fact that Ising couplings on a given sublattice still give rise to a true cat state!

Furthermore, note that this cat state is an eigenstate of the two-point function of
\begin{align}
U Z_{2m-1} U^\dagger
&= Z_{2m-1} \prod_n \exp\left(-\frac{i\pi v_2}{4} Z_{2n-1} Z_{2n+1}\right) \exp\left(\frac{i\pi v_2}{2} X_{2m-1}\right) \prod_n \exp\left(\frac{i\pi v_2}{4} Z_{2n-1} Z_{2n+1}\right) \\
&= Z_{2m-1} \left( \cos(\pi v_2/2) + i \sin(\pi v_2/2) X_{2m-1}
\exp\left(\frac{i\pi v_2}{2} Z_{2m-3} Z_{2m-1}\right) \exp\left(\frac{i\pi v_2}{2} Z_{2m-1} Z_{2m+1}\right)  \right) \\
&= c Z_{2m-1} - s Y_{2m-1} \left( c + i s Z_{2m-3} Z_{2m-1} \right) \left( c + i s Z_{2m-1} Z_{2m-3} \right) \\
&= c Z_{2m-1} - s Y_{2m-1} + O(s^2),
\end{align}
where $c = \cos(\pi v_2/2)$ and $s= \sin(\pi v_2/2)$. This means that the Ising order parameter has simply been rotated, which can be undone by applying a field.

\section{Preparation of other long-range entangled states}

Here we present other states of interest which can be prepared with our protocol. To make this self-contained, we will discuss our first example in detail. Since the other examples work analogously, we will simply quote the result for the stabilizers in those cases.

\subsection{Xu-Moore model (subsystem symmetry breaking)}

Consider the square lattice with the following two sublattices (as in the main text, $A$ is red, $B$ is blue):
\begin{center}
\begin{tikzpicture}
\foreach \y in {0,1,2} {
    \draw[-] (-0.5,\y) -- (4.5,\y);
};
\foreach \x in {0,1,2,3,4} {
    \draw[-] (\x,-0.5) -- (\x,2.5);
};
\foreach \x in {0,2,4}{
    \filldraw[red] (\x,0) circle (2pt);
    \filldraw[blue] (\x,1) circle (2pt);
    \filldraw[red] (\x,2) circle (2pt);
};
\foreach \x in {0,2}{
    \filldraw[blue] (\x+1,0) circle (2pt);
    \filldraw[red] (\x+1,1) circle (2pt);
    \filldraw[blue] (\x+1,2) circle (2pt);
};
\end{tikzpicture}
\end{center}
We start with a product state $|+\rangle^{\otimes (N_A+N_B)}$ and time-evolve with the nearest-neighbor Ising Ham $H = J\sum_{\langle n,m\rangle} Z_n Z_m$ for a time $t_\textrm{SPT} = \frac{\pi}{4|J|}$ (for simplicity we now set $J=1$); the resulting state is a cluster state on the above graph, with stabilizer $X_v \prod_{v' \in \partial v} Z_{v'} = 1$ (where $v' \in \partial v$ denotes that $v$ and $v'$ are nearest neighbors). Upon measuring the $A$ sublattice in the $X$ basis, we obtain the Xu-Moore \cite{Xu04} state on the $B$ sublattice with $\prod_{v \in \square} Z_v = \pm 1$, depending on the measurement outcome. Note that this state corresponds to spontaneous breaking of subsystem symmetries independent of measurement outcome. Acting with judiciously chosen $\prod X_v$ can make it into the homogeneous Xu-Moore state with $\prod_{v\in \square} Z_{v} = +1$.

Despite not having topological order, this is an interesting state with exotic correlations. In particular, there is long-range order for four-point functions of $Z$ if they form a rectangle on the tilted square lattice, whereas one-, two- and three-point functions will vanish! This is thus an analogue of the Schr\"odinger cat state with infinitely many branches in the wavefunction.

\subsubsection{Rydberg implementation: single-species}

If we use a single species of Rydberg atoms (all targeted for the same Rydberg state), then the Rydberg excited states at different sites will interact with a potential $V(r) = U/r^6$. Hence, the first correction to the cluster stabilizer is at the diagonal distance $r=\sqrt{2}a$, giving a stabilizer
\begin{equation}
\langle X_v \prod_{v' \in \partial v} Z_{v'} \rangle \approx \cos^4 \left( \frac{\pi}{2 \sqrt{2}^6} \right) \approx 0.93.
\end{equation}

However, this can be drastically improved by using local addressing, where we presume we can apply individual pulses on the four following sublattices (here $A$ is subdivided into $A$ and $D$; similarly we split $B$ into $B$ and $C$):
\begin{center}
\begin{tikzpicture}
\foreach \y in {0,1,2,3} {
    \draw[-] (-0.5,\y) -- (5.5,\y);
};
\foreach \x in {0,1,2,3,4,5} {
    \draw[-] (\x,-0.5) -- (\x,3.5);
};
\foreach \x in {0,2,4}{
    \filldraw[red] (\x,3) circle (2pt);
    \filldraw[blue] (\x,2) circle (2pt);
    \filldraw[red] (\x,1) circle (2pt);
    \filldraw[blue] (\x,0) circle (2pt);
};
\foreach \x in {0,2,4}{
    \filldraw[purple] (\x+1,0) circle (2pt);
    \filldraw[black] (\x+1,1) circle (2pt);
    \filldraw[purple] (\x+1,2) circle (2pt);
    \filldraw[black] (\x+1,3) circle (2pt);
};
\node[above left,red] at (0,3) {A};
\node[above left,black] at (1,3) {B};
\node[above left,blue] at (0,2) {C};
\node[above left,purple] at (1,2) {D};
\end{tikzpicture}
\end{center}
(We will eventually measure the A and D atoms after preparing the cluster state.) We can effectively generate spin flips $\mathcal X_\lambda = \prod_{v \in \lambda} X_v$ (for any of the sublattices $\lambda=A,B,C,D$) by time-evolving the $X$-term by $\pi/2$. If $U(t)$ denotes time-evolving with the Ising interaction of the system, let us consider the following combined evolution:
\begin{equation}
\mathcal X_A \mathcal X_C U(\pi/8) \mathcal X_A \mathcal X_D U(\pi/8) \mathcal X_C \mathcal X_D U(\pi/4)
\end{equation}

This effectively generates a $t=\pi/4$ evolution for the Ising couplings connecting $AB$, $AC$, $CD$, $BD$, a $t=0$ evolution for $AD$ and $BC$ couplings, and a $t=\pi/2$ evolution for $AA$, $BB$, $CC$, $DD$ couplings. In particular, this entirely eliminates the diagonal corrections (of length $\sqrt{2}$). The result is a stabilizer
\begin{equation}
\langle X_v \prod_{v' \in \partial v} Z_{v'} \rangle \approx \cos^4 \left( 2 \times \frac{\pi}{2 \times 2^6} \right) \approx 0.9952.
\end{equation}
The extra factor of two arose from these contributions adding up over all time-evolutions. Including further corrections, we get a more precise answer:
\begin{equation}
\langle X_v \prod_{v' \in \partial v} Z_{v'} \rangle \approx \cos^4 \left( 2 \times \frac{\pi}{2 \times 2^6} \right) \times \cos^8 \left( \frac{\pi}{2\sqrt{5}^6} \right) \times \cos^4 \left( 2 \times \frac{\pi}{2\sqrt{8}^6} \right) \approx  0.9945,
\end{equation}
confirming that the result is barely changed by including further corrections.

\subsubsection{Rydberg implementation: dual-species}

Let us now instead consider a dual-species implementation, where the same-sublattice ($AA$ and $BB$) couplings can be made negligible. Then the first correction is at a distance $r=\sqrt{5}a$, leading to the cluster stabilizer:
\begin{equation}
\langle X_v \prod_{v' \in \partial v} Z_{v'} \rangle \approx \cos^8 \left( \frac{\pi}{2\sqrt{5}^6} \right) \times \cos^4 \left( \frac{\pi}{2\times 3^6} \right) \approx 0.9994. 
\end{equation}

\subsection{2D color code}

In the main text we discussed how to obtain the toric code on the honeycomb lattice (which required putting qubits on the sites and bonds of the honeycomb lattice). Similarly, we can create the color code \cite{Bombin06} on the honeycomb lattice, which requires putting qubits on the sites of the dice lattice. The resulting stabilizers for the dual-species set-up are:
\begin{align}
\langle X_v \prod_{b\in v} Z_b \rangle &\approx  
 \cos^6 \left( \frac{\pi}{2 \times 2^6} \right) \times
\cos^6 \left( \frac{\pi}{2 \times \sqrt{7}^6} \right) \times \cdots \approx 0.998.\\
\langle B_p \rangle &\approx  \cos^{18} \left( \frac{\pi}{2 \times 2^6} \right) 
\times \cos^{36} \left( \frac{\pi}{2 \times \sqrt{7}^6} \right) \times \cdots \approx 0.994.
\end{align}

\subsection{3D toric code on the diamond lattice}

\begin{figure}
    \centering
    \includegraphics{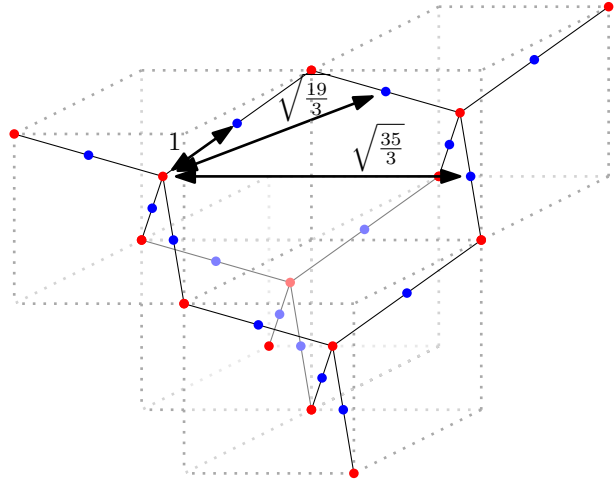}
    \caption{To obtain the toric code on the diamond lattice, we create the cluster state on the above lattice and then measure the $A$ (red) sublattice.}
    \label{fig:diamond}
\end{figure}

Here we consider the 3D toric code \cite{Hamma05}. Instead of placing it on the usual cubic lattice, we find excellent results for the diamond lattice. In particular, consider the $A$ sublattice to be the vertices of the diamond lattice, and the $B$ sublattice the bonds (see Fig.~\ref{fig:diamond}). In the dual-species set-up, we then find:
\begin{align}
\langle X_v \prod_{b\in \textrm{tet}} Z_b \rangle &\approx \cos^{12} \left( \frac{\pi}{2 \cdot \sqrt{19/3}^6} \right) \times  \cos^{24}\left(  \frac{\pi}{2 \cdot \sqrt{35/3}^6} \right) \times \cos^{24}\left(  \frac{\pi}{2 \cdot \sqrt{51/3}^6} \right) \times \cdots \approx 0.9998\\ 
\langle B_p\rangle &\approx 
\cos^{18} \left( \frac{\pi}{2 \cdot \sqrt{19/3}^6} \right) \times  \cos^{36}\left(  \frac{\pi}{2 \cdot \sqrt{35/3}^6} \right) \times \cos^{36}\left(  \frac{\pi}{2 \cdot \sqrt{51/3}^6} \right) \times \cdots \approx 0.9996
\end{align}

\begin{figure}
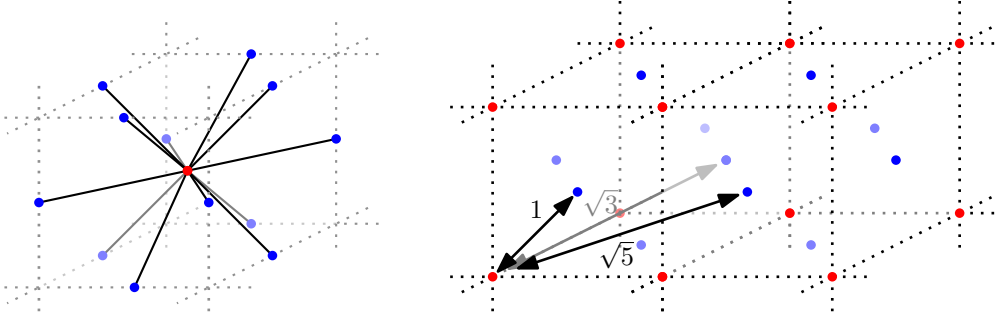

    \centering
    \includegraphics{Xcubeoriginal.pdf}
    \hspace{20pt}
    \includegraphics[scale=1]{Xcubedistance.pdf}
    \caption{We show the connectivity of the cluster state used to prepare the X-cube model. The $A$ sites at the center of the cube are connected to the $B$ sites situated at the twelve edges surrounding the cube.  Together, they form the FCC lattice.}
    \label{fig:Xcube_cluster}
\end{figure}

\subsection{X-cube model}
To realize the X-cube model \cite{Vijay16}, we take the $A$ and $B$ sublattices to be the vertices and the faces centers of the FCC lattice, respectively (see Fig.~\ref{fig:Xcube_cluster}. Defining $A_c = \prod_{b\in \textrm{cube}} Z_b$, we obtain for the dual-species set-up:
\begin{align}
\langle X_v A_c \rangle &\approx \cos^{24} \left( \frac{\pi}{2 \cdot \sqrt{3}^6} \right)  \times \cos^{24}\left(  \frac{\pi}{2 \cdot \sqrt{5}^6} \right)\approx 0.957\\
\langle B_p\rangle &\approx 
\cos^{32} \left( \frac{\pi}{2\cdot \sqrt{3}^6} \right) \times \cos^{32}\left(  \frac{\pi}{2 \cdot \sqrt{5}^6} \right) \approx 0.944
\end{align}

\section{Details for preparing $S_3$ topological order \label{app:S3}}

We first outline formally the preparation of $S_3$ topological order. We then discuss the physical implementation in Rydberg atom arrays.

\subsection{Formal preparation}

Our preparation consists of two steps.  Preparing the $\ZZ_3$ toric code by gauging a product state with $\ZZ_3$ symmetry, and further gauging the $\ZZ_2$ charge conjugation symmetry to obtain the $S_3$ topological order.

\begin{figure}[h!]
    \centering
    \includegraphics{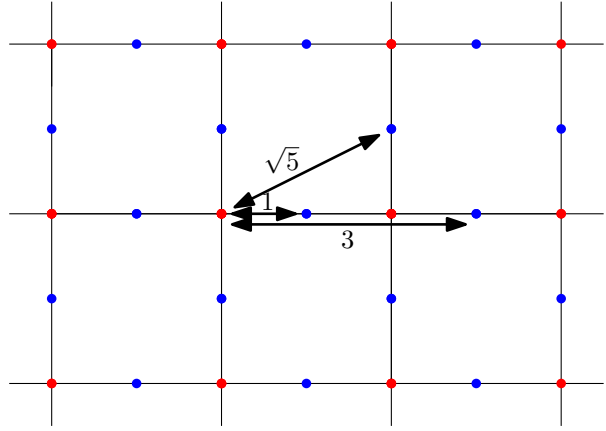}
    \caption{The Lieb lattice.}
    \label{fig:lieb}
\end{figure}

The lattice we will use in the first step is the so-called Lieb lattice with the $A,B$ sublattice on the vertices and bonds of the square lattice, respectively (see Fig.~\ref{fig:lieb}). We denote
\begin{align}
\cX&=\begin{pmatrix}
0 &0 &1\\
1&0&0\\
0&1&0
\end{pmatrix} &
\cZ &= \begin{pmatrix}
1 &0 &0\\
0&\omega&0\\
0&0&\bar \omega
\end{pmatrix} &
\cC &= \begin{pmatrix}
1 &0 &0\\
0&0&1\\
0&1&0
\end{pmatrix}
\end{align}
to be the shift, clock, and charge conjugation of qutrits respectively. Here $\omega = e^{2\pi i/3}$.

We initialized a qutrit in the state $\ket{+} =\frac{\ket{0}+ \ket{1}+\ket{2}}{\sqrt{3}}$, which is an eigenstate of $\cX$. Thus the state is uniquely specified by stabilizers $\cX$ for every site. Next, we perform an evolution that creates the $\ZZ_3$ cluster state on this lattice. This cluster state can be viewed as an SPT protected by a $\ZZ_3$ global symmetry and a $\ZZ_3$ 1-form symmetry. The unitary is given by
\begin{align}
    U_{AB} = \prod_{\inner{a,b}} C\mathcal Z_{ab}
    \label{eq:UAB}
\end{align}
where $\inner{a,b}$ denotes nearest neighbors for $a\in A$ and $b\in B$ and $C\cZ = \omega^{n_a n_b}$ is the controlled$-\cZ$ gate. Conjugating by this unitary, the stabilizers are given by
\begin{align}
    \raisebox{-.5\height}{ \includegraphics[]{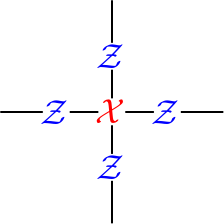} } &&  \raisebox{-.5\height}{ \includegraphics[]{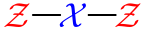} } && \raisebox{-.5\height}{ \includegraphics[]{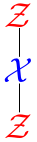} }  
\end{align}
Next we perform a measurement on the $A$ sublattice in the $\mathcal X$ basis. In the ideal case, when all measurement outcomes are $\ket{+}$, we obtain the ground state of the $\ZZ_3$ toric code, given by stabilizers
\begin{align}
    \raisebox{-.5\height}{ \includegraphics[]{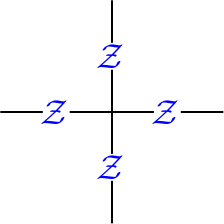} } &&  \raisebox{-.5\height}{ \includegraphics[]{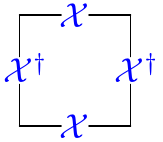} } &&
\end{align}
The other measurement outcomes which are eigenstates of $\mathcal X$ signify the presence of $e$ and $\bar e$ anyons of the toric code, which can be paired up and annihilated via local unitaries.

The toric code wavefunction respects a global charge conjugation symmetry $\prod_B \cC$. In order to gauge this symmetry, we will add two additional sublattices $C$ and $D$ as shown consisting of qubits initialized with stabilizers $X$. We will use $C$ to perform a symmetry transformation so that the charge conjugation symmetry acts as $\prod_C X$. From this, we can then gauge this $\ZZ_2$ symmetry by coupling the $C$ and $D$ lattices using the $\ZZ_2$ cluster state entangler and measuring the $C$ sublattice.

\begin{figure}[h!]
    \centering
    \includegraphics{2DsquareS3.pdf}
    \caption{Sublattices $C$ (red) and $D$ (purple) which form a shifted Lieb lattice (dotted lines) are added. Each site hosts a qubit.}
    \label{fig:my_label}
\end{figure}

The symmetry transformation we will perform will exchange the following symmetries
\begin{align}
    \prod_B X  \leftrightarrow \prod_B X \prod_C \cC
\end{align}
The latter acts enriches the $\ZZ_3$ toric code by charge conjugating both $e$ and $m$ anyons. The unitary that achieves this transformation is given by
\begin{align}
    U_{BC} = \prod_{\inner{b,c}} C\mathcal C_{cb}
    \label{eq:UBC}
\end{align}
where $\inner{bc}$ denotes nearest neighbor sites of $b\in B$ and $c \in C$, and $C\cC_{cb}$ is the controlled-$\cC$ gate where $c$ is the control. The stabilizers are now
\begin{align}
        \raisebox{-.5\height}{ \includegraphics[]{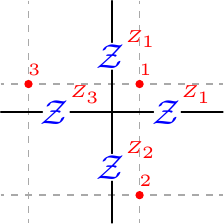} } &&  \raisebox{-.5\height}{ \includegraphics[]{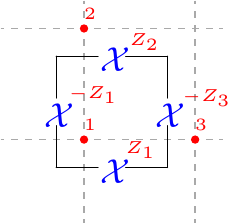} } &&
        \raisebox{-.5\height}{ \includegraphics[]{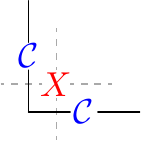} } &&
\end{align}

We then couple the $C$ and $D$ sublattices with a $\ZZ_2$ cluster state. This is achieved via
\begin{align}
    U_{CD} = \prod_{\inner{c,d}}CZ_{cd}
    \label{eq:UCD}
\end{align}

At this stage the stabilizers of the wavefunction are given by
\begin{align}
        \raisebox{-.5\height}{ \includegraphics[]{AvZ3TCcouple.pdf} } &&  \raisebox{-.5\height}{ \includegraphics[]{BpZ3TCcoupled.pdf} } &&
        \raisebox{-.5\height}{ \includegraphics[]{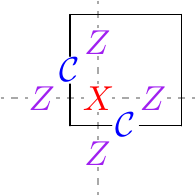} } &&
         \raisebox{-.5\height}{ \includegraphics[]{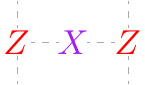} } &&
          \raisebox{-.5\height}{ \includegraphics[]{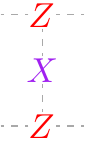} } &&
\end{align}

Now, we perform a measurement on the $C$ sublattice. It is clear that the following vertex and plaquette terms constructed from the order two stabilizers are stabilizers of the final model:
\begin{align}
         \raisebox{-.5\height}{ \includegraphics[]{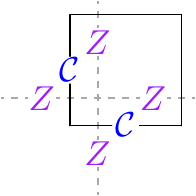} } &&
          \raisebox{-.5\height}{ \includegraphics[]{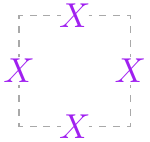} } &&
\end{align}
On the other hand, the order three stabilizers do not commute. Let us call the first $\ZZ_3$ stabilizer $A_v$. We note that $\frac{1+A_v + A_v^\dagger}{3}$ is a projector for the wavefunction that commutes with the measurement $X$ on the $C$ sublattice. Therefore, we can write down the resulting projector after the measurement. Let us denote the number operator of the four qutrits according to the cardinals $n_N$, $n_E$, $n_W$, $n_S$. Since the operator is diagonal, it acts on the basis states as

\begin{align}
    A_v = \omega^{n_N Z_1 + n_E Z_1 + n_S Z_2 + n_W Z_3}
\end{align}
Therefore,
\begin{align}
    A_v +A_v^\dagger= 2 \cos\left [ \frac{2\pi}{3} Z_1 (n_N + n_E + n_S Z_1Z_2 + n_WZ_1Z_3)\right ]
\end{align}
where we have pulled out a common factor $Z_1$. Since this operator has eigenvalues $\pm 1$, it can be removed from the cosine. We are left with
\begin{align}
    A_v +A_v^\dagger&= 2 \cos\left [ \frac{2\pi}{3} Z_1 (n_N  + n_E + n_SZ_1Z_2 + n_W Z_1Z_3)\right ]\\
    &= \omega^{n_N  + n_E  + n_SZ_1Z_2 + n_WZ_1Z_3} + h.c.\\
    &= \raisebox{-.5\height}{ \includegraphics[]{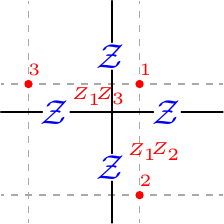} } +h.c.
\end{align}
Therefore, combining with the order two stabilizers, we find that the projector after the measurement is
\begin{align}
\frac{1}{3}\left [1 + \left (\raisebox{-.5\height}{ \includegraphics[]{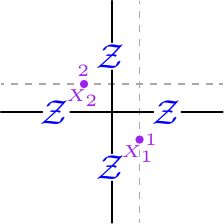} } + h.c.\right) \right]
\end{align}

Similarly, the projector that results from the order three plaquette term after the measurement is
\begin{align}
\frac{1}{3}\left [1 + \left (\raisebox{-.5\height}{ \includegraphics[]{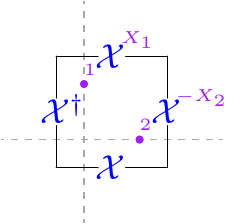} } + h.c.\right) \right]
\end{align}

To conclude, our final wavefunction is not described by a stabilizer code, but instead by the following commuting projectors
\begin{align}
 \frac{1}{3}\left [1 + \left (\raisebox{-.5\height}{ \includegraphics[]{AvZ3TCcouple3.pdf} } + h.c.\right) \right] &&
  \frac{1}{2}\left [1+ \raisebox{-.5\height}{ \includegraphics[]{Avs.pdf} } \right] \\
  \frac{1}{3}\left [1 + \left (\raisebox{-.5\height}{ \includegraphics[]{BpZ3TCcoupled2.pdf} } + h.c.\right) \right]&&
  \frac{1}{2}\left [1+ \raisebox{-.5\height}{ \includegraphics[]{Bps.pdf} }\right] 
\end{align}

\subsection{Relation to the quantum double model $\mathcal D(S_3)$}
To make a connection to the quantum double model, we shift the two square lattices so that they coincide. Each edge now has a Hilbert space of dimension six composed of a qutrit and a qubit. We will show that the projectors can be mapped exactly to those of the quantum double model defined on the square lattice.

To make the explicit connection, we will need to do the following transformations (which mutually commute)
\begin{enumerate}
    \item Swap $\mathcal Z$ and $\mathcal X$ on all qutrits;
    \item Swap $Z$ and $X$ on all qubits;
    \item Apply charge conjugation $\mathcal C$ in the following pattern.
\end{enumerate}

\begin{figure}[h!]
    \centering
    \includegraphics{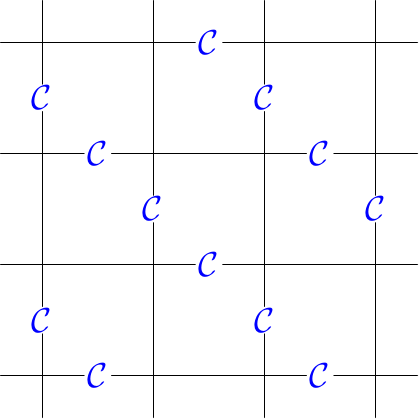}
    \caption{Pattern of $\cC$ operators}
    \label{fig:Cpattern}
\end{figure}

After the above transformations. The projectors take the following form.

\begin{align}
 \frac{1}{3}\left [1 + \left (\raisebox{-.5\height}{ \includegraphics[]{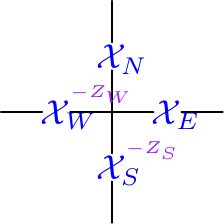} } + h.c.\right) \right] &&
  \frac{1}{2}\left [1+ \raisebox{-.5\height}{ \includegraphics[]{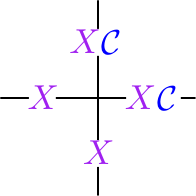} } \right] \\
  \frac{1}{3}\left [1 + \left (\raisebox{-.5\height}{ \includegraphics[]{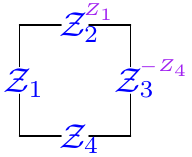} } + h.c.\right) \right]&&
  \frac{1}{2}\left [1+ \raisebox{-.5\height}{ \includegraphics[]{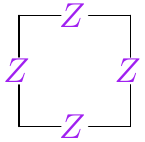} }\right] 
\end{align}

We claim that these are exactly the projectors of the quantum double model of $S_3$. The projectors for the quantum double model are the following vertex and plaquette terms\cite{Kitaev_2003}
\begin{align}
    \bs A_v &= \frac{1}{|G|}\sum_{g\in G}  \raisebox{-0.5\height}{\includegraphics[scale=1]{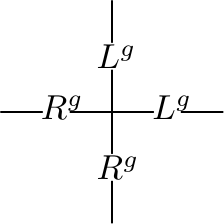}}\\
    \bs B_p &=\sum_{g_1,g_2,g_3,g_4 \in G} \delta_{1,g_1g_2 g_3^{-1} g_4^{-1}} \Ket{\raisebox{-0.5\height}{\includegraphics[scale=1]{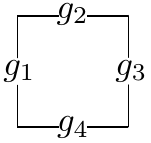}}}\Bra{\raisebox{-0.5\height}{\includegraphics[scale=1]{BpQD.pdf}}}
\end{align}
where $L^g$ and $R^g$ are left and right multiplication operators
\begin{align}
L^g \ket{h} &= \ket{gh} & R^g\ket{h} = \ket{h\bar g}
\end{align}
Let us represent the group $S_3$ using the generators $r,s$ that satisfy $r^3=s^2=srsr=1$. In the regular representation, the left and right multiplications of $S_3$ are $6 \times 6$ matrices, which can be explicitly written in terms of a qubit and a qutrit as
\begin{align}
    L^r&=I \otimes \cX  & L^s &= X \otimes \cC\\
    R^r&= \cX^\dagger \oplus \cX = \cX^{-Z} & R^s &=X \otimes I
\end{align}
Therefore, the product of our two vertex projectors is exactly $\boldsymbol A_v$. Similarly, one can show that the product of our plaquette projectors is exactly $\boldsymbol B_p$ (see appendix A of Ref. \cite{TJV2} for a proof).

\subsection{Implementation with Rydberg atoms}

To realize a qutrit on the $A$ and $B$ sublattices, we endow each site with two Rydberg atoms, which we call atoms $1$ and $2$. Due to the proximity of the atoms, the state where both atoms are excited is prohibited, giving an effective three-level system. We label the empty state as $\ket{0}$ and the state where atom 1 or 2 is excited as $\ket{1}$ or $\ket{2}$ respectively. On the other hand, the $C$ and $D$ sublattice will be the usual Rydberg two-level system: $\ket{0}$ for the ground state and $\ket{1}$ for the excited state.

Assuming local addressing, we are able to perform the following pulses for the qutrits:
\begin{align}
    X_1 &= \begin{pmatrix}
    0&1&0\\
    1&0&0\\
    0&0&0
    \end{pmatrix}, &    Z_1 &= \begin{pmatrix}
    1&0&0\\
    0&-1&0\\
    0&0&0
    \end{pmatrix}, & X_2 &= \begin{pmatrix}
    0&0&1\\
    0&0&0\\
    1&0&0
    \end{pmatrix}, &  Z_2 &= \begin{pmatrix}
    1&0&0\\
    0&0&0\\
    0&0&-1
    \end{pmatrix}    
\end{align}
which are the result of projecting the spin rotation operators into the qutrit subspace. These four generators are enough to generate arbitrary $SU(3)$ rotations.

The preparation procedure can be summarized as follows:
\begin{enumerate}
    \item Load the $A$ and $B$ sublattices and perform the $\ZZ_3$ fourier transform;
    \item Implement $U_{AB}$ Eq.~\eqref{eq:UAB};
    \item Perform the inverse $\ZZ_3$ fourier transform on the $A$ sublattice and measure the occupancy ($\ZZ_3$ toric code is obtained at this step);
    \item Perform single site rotations to pair up charges in the $\ZZ_3$ toric code;
    \item Load the $C$ sublattice and perform Hadamard;
    \item Implement $U_{BC}$ Eq.~\eqref{eq:UBC};
    \item Load the $D$ sublattice and perform Hadamard;
    \item Implement $U_{CD}$  Eq.~\eqref{eq:UCD};
    \item Perform Hadamard on the $C$ sublattice and measure. ($S_3$ topological order is obtained at this step);
    \item (optional) Perform single site rotations to pair up charges in the $S_3$ topological order.
\end{enumerate}

We now discuss how to implement $U_{AB}$, $U_{BC}$, and $U_{CD}$.

\subsubsection{$U_{AB}$}
To create the $\ZZ_3$ cluster state, we would like to generate the unitary $U_{AB}$ in Eq.~\eqref{eq:UAB}.

The innate two-body interaction has dominated by the nearest neighbor sites between sublattices $A$ and $B$, and is given by
\begin{align}
       H_{AB}(U,U') = \sum_{\inner{ab}} \left[ U \left( n_{1,a} n_{1,b} +  n_{2,a} n_{2,b} \right) +  U' \left( n_{1,a} n_{2,b}+  n_{2,a} n_{1,b} \right) \right], \label{eq:H_AB}
\end{align}
where we presume $1\leftrightarrow 1$ and $2\leftrightarrow 2$ are equidistant, as are $1\leftrightarrow2 $ and $2\leftrightarrow1$. (One way of achieving this is by having the two atoms separated in the third direction.) Here, $\inner{a,b}$ denote nearest neighbors, and $n_1$ and $n_2$ denotes the occupancy of atoms $1$ and $2$, respectively. Note that by conjugating the number operator $n_i$ with a $\frac{\pi}{2}$ pulse of $X_j$ for $j=1,2$ gives
\begin{align}
    e^{\frac{\pi i}{2} X_j} n_j e^{-\frac{\pi i }{2} X_j} = 1-n_1-n_2.
\end{align}
One can show that by successive applications of this substitution, as well as the freedom of tuning single-site chemical potentials, one can generate an effective evolution where $U'=-U$. In fact, if $U' > U$ in Eq.~\eqref{eq:H_AB}, the effective evolution will have $U = -U' > 0$. If $U'<U$, one can simply exchange the $1\leftrightarrow 2$ labels on the $A$ sublattice to reduce back to $U'>U$. In conclusion, we thus obtain the $C \mathcal Z$ gate
$U_{AB}$ via
\begin{align}
    U_{AB} = e^{-4\pi i/(3U) H(U,-U)}.
\end{align}

\subsubsection{$U_{BC}$}
To implement $U_{BC}$ in Eq.~\eqref{eq:UBC}, we first perform a basis transformation on the $B$ sublattice such that $\cC$ is diagonal. A choice is
\begin{align}
    U&= e^{\frac{\pi i}{4}X_1}e^{\frac{\pi i}{4}Z_1}e^{\frac{\pi i}{2}X_2}e^{-\frac{\pi i}{2}Z_2}\\
    \tilde \cC \equiv U\cC U^\dagger &= \begin{pmatrix} -1&0&0\\
    0&1&0\\ 0&0&1\end{pmatrix}
\end{align}

The leading order interactions between the $B$ and $C$ sublattices are of the form
\begin{align}
       H_{BC} = \sum_{\inner{bc}}n_{1,b} n_{c}  + n_{2,b} n_{c} 
\end{align}
Since
\begin{align}
    e^{\pi i n_c(n_{1,b}+n_{2,b})}= \textrm{diag}(1,1,1,1,-1,-1)= Z_c  \cdot C\tilde \cC_{c,b}
\end{align}
where $C\tilde \cC$ is the controlled $\tilde \cC$ gate. We have that
\begin{align}
    e^{\pi i H_{BC}}= \prod_{\inner{b,c}}  C\tilde \cC_{c,b}
\end{align}
where we used the property that for a fixed $c\in C$, there are only two nearest neighbors $b \in B$, thus cancelling away $Z_c$. To conclude,
\begin{align}
    U_{BC} = \prod_{\inner{b,c}} C\cC_{c,b} = U e^{\pi i H_{BC}} U^\dagger
\end{align}

\subsubsection{$U_{CD}$}
$U_{CD}$ in Eq.~\eqref{eq:UBC} is the $\ZZ_2$ cluster state entangler between the $C$ and $D$ sublattices, so we can implement this via the ordinary time evolution. However, we also have to ensure that there is no net evolution between the pairs $BC$ and $BD$. We can do this by flipping all $C$ and $D$ qubits halfway through the evolution. That is,
\begin{align}
    U_{CD} = \left (\prod_{C,D} X\right) e^{\frac{\pi i}{2} H} \left (\prod_{C,D} X\right) e^{\frac{\pi i}{2} H}
\end{align}

\end{document}